\documentclass{jjap3}
\usepackage{newtxtext,newtxmath}

\usepackage{lastpage}
\usepackage{array}
\usepackage[dvipdfmx]{graphicx}
\usepackage{color}
\usepackage {diagbox} 
\usepackage{slashbox}
\usepackage{multirow}
\usepackage{time}
\usepackage{bm}

\renewcommand*{\epsilon}{\varepsilon}

\definecolor{refkey}{rgb}{0.9451,0.2706,0.4941}
\definecolor{labelkey}{rgb}{0.9451,0.2706,0.4941}

\title{Reactive Molecular Dynamics Simulation on DNA Double Strand Breaks Induced by Hydrogen Elimination}
\author{Hiroaki Nakamura$^{1,2}$\thanks{E-mail: hnakamura@nifs.ac.jp},
Kento Ishiguro$^2$,
Ayako Nakata$^3$, 
Shunsuke Usami$^{1,4}$, 
Seiki Saito$^5$ 
and Susumu Fujiwara$^{6}$}
\inst{
$^{1}$National Institute for Fusion Science, 322-6 Oroshi-cho, Toki, Gifu 509-5292, Japan\\
$^{2}$Nagoya University, Furo-cho, Chikusa-ku, Nagoya 464-8601, Japan\\
$^{3}$National Institute for Materials Science (NIMS), 1-1 Namiki, Tsukuba, Ibaraki 305-0044, Japan\\
$^{4}$The University of Tokyo, Tokyo 113-8654, Japan\\
$^{5}$Yamagata University, 4-3-16 Jonan, Yonezawa, Yamagata 992-8510, Japan\\
$^{6}$Kyoto Institute of Technology, Matsugasaki, Sakyo-ku, Kyoto 606-8585, Japan
}
\abst{
We propose a scar model to reproduce how double-strand breaks occur in the telomeric DNA damaged by the effect of the $\beta$-decay of tritium to helium.
In this scar model, the two hydrogens bonded to the 5$^\prime$ carbon connecting the pentasaccharides and phosphate are removed. 
Molecular dynamics simulations using the reactive force field are carried out for 10 cases for the telomeric DNA consisting of 16 base pairs (32 nucleotides). 
It results in double-strand breaks (DSBs) being observed for structures with more than 24 scars. 
For 16 scar cases, only single-strand breaks (SSB) are observed. 
Moreover, in the case of $\left\{16, 0 \right\}$ and $\left\{ 0, 16 \right\},$ where only one of the strands had scars, SSB occurs only in the scarred strand. 
Secondly, in the $\left\{16, 8 \right\}$ and $\left\{ 8, 16 \right\}$ cases, DSBs occurs. 
Therefore, we conclude that the following conditions are necessary for DSBs:
(i) Scars must occur on both the L and R strands. 
(ii) A large number of scars (24 or more) must occur in close proximity to each other.
 }

\begin{document}
\maketitle

\section{Introduction}
Molecular dynamics simulations using reactive force fields were developed by Brenner for  Carbon nanotubes\cite{02Brenner}. 
This force field can handle the bonding and breaking of multiple atoms for a limited number of atom elements such as carbon, hydrogen, and oxygen. 
It also defines a measure of bond order and can distinguish the order of covalent bonds. 
This reactive force field has been improved, and now a reactive force field that can handle more elements, including biopolymers and metals, has been proposed.
Our group has performed calculations using Brenner's reactive force field to treat the wall of a fusion reactor\cite{06Ito, 15Ito, 11Naka, 23Naka}. 
Recently, we have attempted to treat the behavior of DNA\cite{20Naka,22Hatano, 19Fujiwara, 19Li} by using this experience.

Several years ago, we started using molecular dynamics (MD) simulations\cite{Allen, Frenkel} to assess tritium-induced DNA damage,\cite{17Tanabe,16Tritium, 09Mull} including transmutation effects.\cite{20Naka, 22Hatano}
Simultaneously with the simulations, we also started DNA damage experiments\cite{18Hatano,22Hatano} that were complementary to the simulations.

In our first work of DNA,\cite{20Naka} we focused on the telomere structure at the end of DNA and performed MD simulations, assuming that tritium (T) is taken into the cell for some reason and that this tritium replaces hydrogen (H) in the guanines in telomeres and eventually $\beta$-decays to helium (He).
Here, we used, instead of the reactive force field, only a conventional force field CHARMM36,\cite{CHARMM36DNA} which can calculate the optimal DNA shape in solvent but cannot handle covalent bond cleavage or synthesis.
This was because, at the time, we could not find a reactive force field that could handle DNA.
Thus, using the CHARMM36 force field, we were able to reproduce the collapse of the double helix structure as the hydrogen in the guanine of the telomere decays to helium, weakening the hydrogen bonds between the DNA double strands.
As a result, we evaluated quantitatively the rate at which the double strands unwind as a consequence of $\beta$-decay of T in guanine to He.
However, the CHARMM36 force field used in this study cannot accommodate the most intriguing phenomenon, namely double-strand breaks of DNA.

As time has passed since our previous simulations and we were able to find some candidates of reactive force field that can handle DNA, we take up the challenge of double-strand breaks of DNA, which we had been focusing on for some time.
Specifically, we calculate the case in which the 5$^\prime$  hydrogens connecting the pentasaccharides and phosphate, which is thought to be most likely to undergo tritium substitution within the telomere, is desorbed.
This paper describes the molecular dynamics (MD) simulations, in this case,  on double-strand breaks (DSBs)  performed using a reactive force field (ReaxFF).

\section{Molecular Dynamics Simulation Model}
\subsection{Telomeric DNA}
As in our previous work\cite{20Naka}, we pick up telomeres which are structures at the ends of eukaryotic chromosomes that consist of repeats of specific base sequences (telomeric DNA).

The reason for choosing telomeric DNA as the simulation model are the same as in the previous study, as follows:
The shortening of telomeres replication in human cells has an important role in cellular senescence\cite{99Greider}. 
Therefore, we consider that the human cell damage by tritium depends on the damage of the telomeric DNA.

\begin{figure} \centering
\includegraphics[width=3cm]{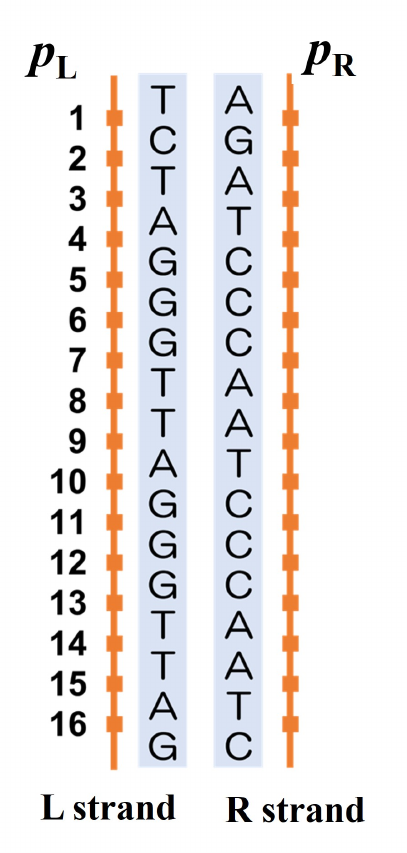}
\caption{ 
The base sequence of the DNA.  
The four types of bases are adenine A, guanine G, cytosine C, and thymine T. 
The orange squares per strand denote the phosphorous atoms, which are numbered on the L-strand with $p_\textrm{L}$ and on the R-strand with $p_\textrm{R}.$  
The side with the sequence $\left\{\right.$TCTAGGGTTAGGGGTTAG$\left\}\right.$ is denoted as L-strand, and the side with $\left\{\right.$AGATCCCAATCCCAAT$\left\}\right.$  as R-strand. 
}
\label{fig0010}
\end{figure}

The structure of the telomeric DNA is obtained by removing TRF2 protein from the TRF2-Dbd-DNA complex, PDB ID  of which is 3SJM\cite{3SJM}.
The telomeric DNA has 17 base pairs, d(TCTAGGGTTAGGGTTAG), which consists of 1,078 atoms as shown in Fig. \ref{fig0010}.
Here, the four types of bases are adenine A, guanine G, cytosine C, and thymine T. 
To distinguish between the two strands, the side with the sequence $\left\{\right.$TCTAGGGTTAGGGGTTAG$\left.\right\}$ is denoted as L-strand, and the side with $\left\{\right.$AGATCCCAATCCCAATC$\left.\right\}$ as R-strand. 
In addition, the orange squares in Fig.\ref{fig0010} indicate the phosphorus atoms in the nucleotides with 16 phosphorus atoms per strand.


Since the telomeric DNA used in this simulation is originally negatively charged ($- 32 e$, where the elementary charge $e = 1.602176634 \times 10^{-19}$ C ), sodium ions are added to neutralize the entire system. 
Furthermore, the salt concentration must be set to the concentration of the human body. 
Considering the above, 120 sodium ions (Na$^+$) and 88 chlorine ions (Cl$^-$) are added randomly and placed in the system.
The initial volume $V$ of the simulation box is 100 \AA $\times$ 100 \AA $\times$ 100 \AA. 
We also add 30,773 water molecules into the simulation box.

\begin{figure}[htbp] \centering 
\includegraphics[width=5cm]{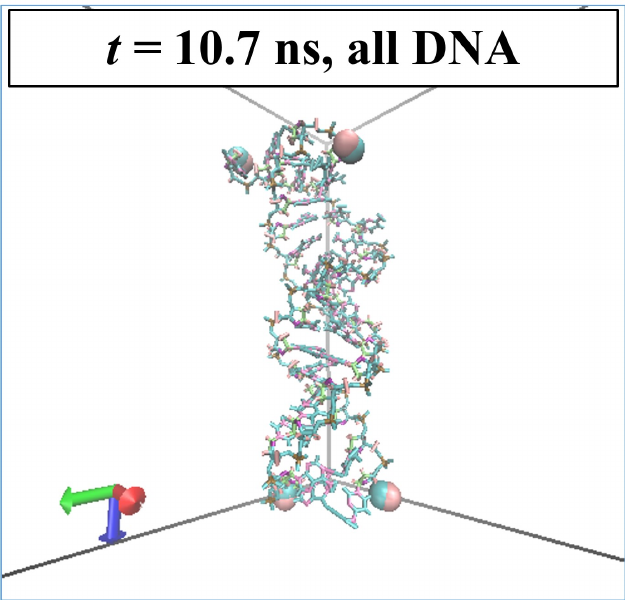}
\hspace{0.8cm}
\includegraphics[width=5cm]{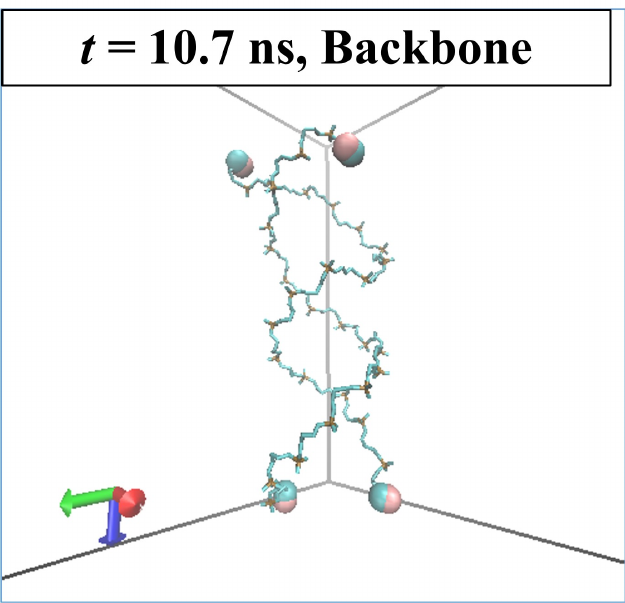}
\caption{
The left figure shows snapshots of the structure of telomeric DNA at $t$ = 10.7 ns. 
All the atoms that make up the telomeric DNA are drawn.
In the right figure, only the main strands are picked up.
In the third process, the four hydroxyl (OH) groups at the ends of the DNA strand  (large balls) are fixed to prevent the DNA from collapsing from the ends.
}
\label{fig0020}
\end{figure}

\begin{figure}[htbp] \centering 
\includegraphics[width=9cm]{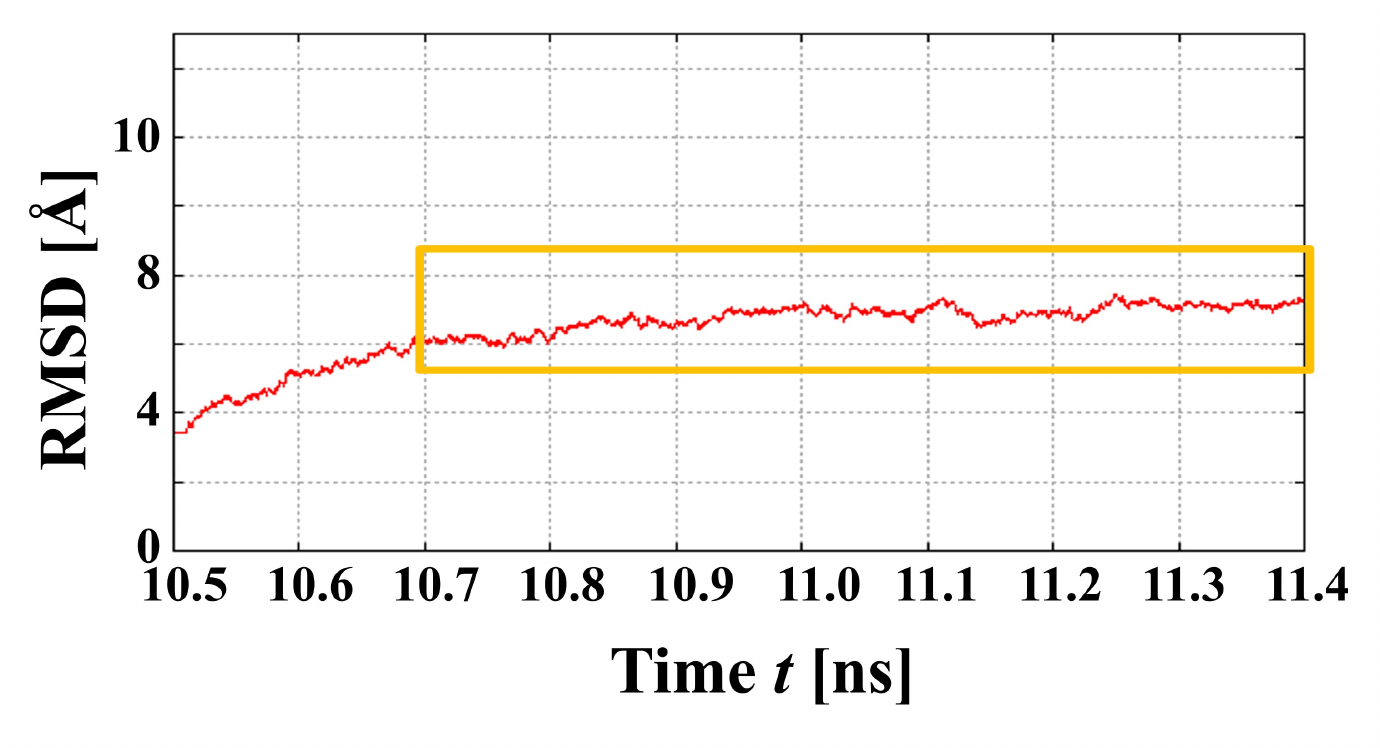}
\caption{
Time evolution of the root mean square deviation (RMSD) of the atoms composing the telomeric DNA molecule.
The RMSD remains almost constant between 10.7 ns and 11.4 ns.
}
\label{fig0030}
\end{figure}

First, we make the steady state of the telomeric DNA with the solvent at 310K.
To achieve the steady state, the following three-step process is used as time $t$ elapses:
\begin{enumerate}
\item{$0\le t \le $ 0.5 ns and $T$ increases from 0 K to 310K  ($NVT$ ensemble, CHARMM force field): }
At $t=0$, we perform the energy minimization of the total system.
Then, using canonical ensemble ($NVT$ ensemble), we increase linearly the reference temperature $T$ from 0 K to 310 K over 0.5 ns while the coordinates of the DNA are ﬁxed.
The volume $V$ of the simulation box is kept to  100 \AA $\times$ 100 \AA $\times$ 100 \AA. 
The periodic boundary condition is adopted in the $x, y,$ and $z$ directions.
In this first process,  CHARMM force field is adopted\cite{CHARMMpot, CHARMM36DNA} using Large-scale Atomic/Molecular Massively Parallel Simulator (LAMMPS) code \cite{LAMMPS, LAMMPSURL} with Langevin thermostat algorithm\cite{Langevin} to control the temperature of the system.
The time step of the MD simulation is 1 fs.
In the first process, the coordinates of all atoms that make up telomeric DNA are fixed. 
The solvent molecules, on the other hand, evolve in time. 
Eventually, the Na$^+$ and Cl$^-$ ions in the solvent that have been interacting with the negative charge of the DNA rearrange themselves, and the solvent reaches a steady state.
We comment here on the difference between the CHARMM force field and the reactive force field\cite{v2,ReaxFF} (ReaxFF).
CHARMM force field was developed to calculate the shape change of a biomolecule (in this case,  telomeric DNA) in a solvent, but it cannot handle covalent bond cleavage or formation of biomolecules.
Therefore, it can be faster than calculations using the ReaxFF, which can handle covalent bond cleavage and formation.

\item{0.5 ns $\le t \le $10.5 ns, $T$ is fixed at 310 K and $P$ is fixed at 1 bar ($NPT$ ensemble, CHARMM force field): }
After the reference temperature $T$ reaches 310 K in the first process, $T$ is maintained at 310 K thereafter. 
Constraints of the telomeric DNA are gradually released.
Moreover, to examine the behavior of DNA under $NPT$ ensemble with 1 bar, the simulation conditions are transferred from $NVT$ ensemble to $NPT$ ensemble with sequential changes in the constraints of the telomeric DNA.
The details of the simulation conditions are explained in turn below.
After $t=$ 0.5 ns, the simulation condition is changed from $NVT$ ensemble to $NPT$ ensemble where the pressure $P$ is fixed at 1 bar.
The total number of particles $N$ is kept in the first process state.
Conversely, the volume $V$ of the simulation box is no longer treated as fixed but variable.
From 0.5 ns to 1.5 ns, the coordinates of all atoms that make up telomeric DNA are still fixed,  as in the first process.
From 1.5 ns to 3.5 ns, the coordinates of atoms that compose the main strand of the telomeric DNA are fixed.
On the other hand, the atoms of the side chains of the telomeric DNA are not fixed and are movable.
From 3.5ns to 10.5 ns, all atoms of telomeric DNA are unfixed and the DNA is free to change its shape.
Thus, we obtain the stable structure of the telomeric DNA and the solvent under the CHARMM force field at 310 K and 1 bar.
\item{10.5 ns $\le t \le$ 11.4  ns, $T$ is fixed at 310 K and $P$ is fixed at 1 bar ($NPT$ ensemble,  ReaxFF):}
Next, MD simulations are performed using ReaxFF, which can handle covalent bond cleavage or formation.
However, simply changing the force field from CHARMM to ReaxFF\cite{Reax18} in the configuration of the molecules obtained in the previous process makes the system unstable and collapse from the ends of the DNA strands.
Therefore, the four hydroxyl (OH) groups at the ends of the DNA strand are fixed to prevent the DNA from unintentionally disintegrating its structure from the ends (See Fig. \ref{fig0020}).
 With this constraint of the four ends, the MD simulations using the ReaxFF are performed under $NPT$ ensemble ($T$ = 310 K, $P$ =1 bar) conditions.
The time step is 0.1 fs, which is smaller than in the case of CHRAMM.
In this way, the stable structure of the telomeric DNA in solvent is created.
In addition, the time evolution of the root mean square deviation (RMSD)\cite{20Naka} is shown in Fig. \ref{fig0030}. 
it can be seen that the RMSD becomes almost stable in the region of 10.7 ns $ \le t \le $ 11.4 ns. 
This confirms that our created telomeric DNA structure is stable. 
 \end{enumerate}
 
 

\subsection{Scar Model}
\subsubsection{Definition of scar model}
\begin{figure}[htbp] 
\includegraphics[width=6cm]{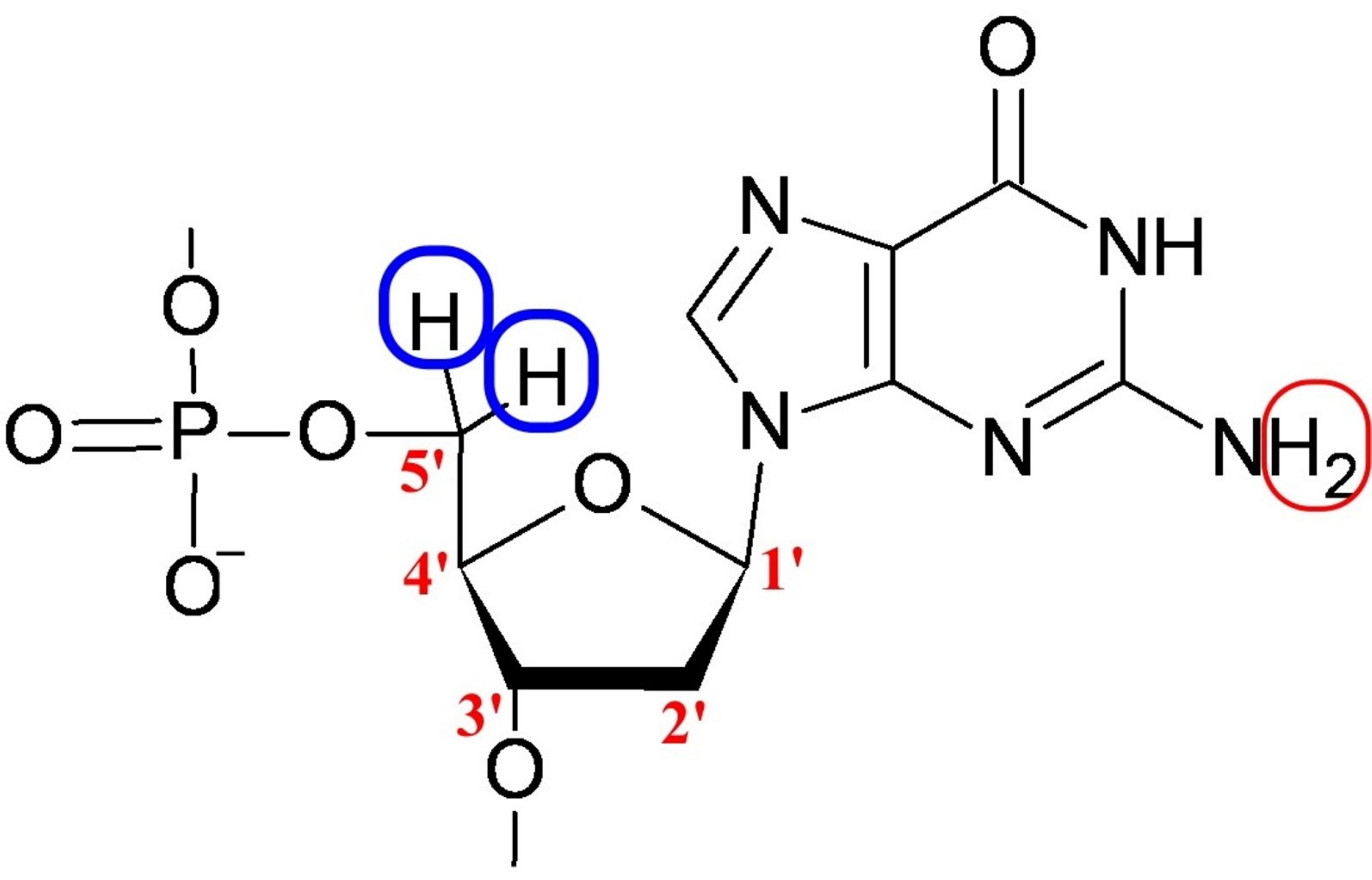}
\caption{
Molecular structural formula of ``scar model.''
This nucleotide is depicted for guanine as a base.
The two hydrogens bonded to the 5$^\prime$ carbon between the pentasaccharide and the phosphate group are highlighted by blue circles.
In the scar model, these two hydrogens are removed.
Incidentally, our previous work\cite{20Naka} employed the model in which the two hydrogens encircled in red in the guanine were removed.
}
\label{fig0040}
\end{figure}

In this simulation, DNA damages are treated as ``the removal of two hydrogens binding to the 5$^{\prime}$ carbon 
between the pentasaccharides and the phosphate group'' as shown in Fig.\ref{fig0040}. 
This hydrogen-removed state will henceforth be referred to as a `scar'. 
This scar model is based on the fact that molecular dynamics simulation by Tsuchida et al.\cite{Tsuchida} shows that atoms in the solvent collide with these hydrogens with the highest frequency. 
In this case, if tritiums are present in the solvent, the probability of hydrogen substitution between tritium in the solvent and hydrogen in the DNA is expected to be highest for this 5$^{\prime}$  bound hydrogen. 
When this substitution occurs, tritium bound to the 5$^{\prime}$   carbon will proceed to $\beta$-decay, decaying to helium with a half-life of about 12.3 years.\cite{17Tanabe} 
Unlike hydrogen isotopes, the helium produced in this reaction is not covalently bound to the 5$^{\prime}$ carbon, so it eventually diffuses into the solvent and does not remain in the DNA. 
This is the phenomenon assumed in the scar model.
The model has the advantage that it can be applied not only to the disintegration effect of tritium $\beta$-decay, as originally intended, but also to cases where hydrogen desorption occurs by direct or indirect action.

\begin{figure}[htbp] \centering 
\includegraphics[width=5cm]{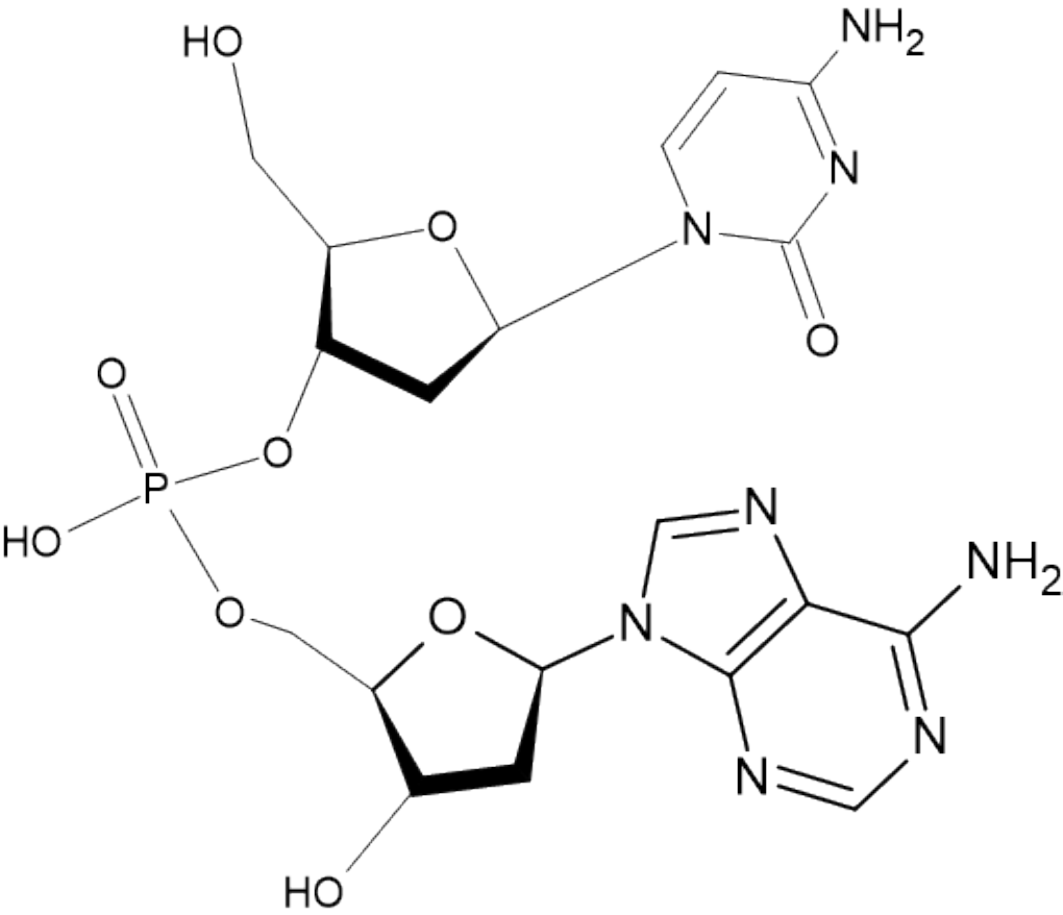}
\hspace{1cm}
\includegraphics[width=5cm]{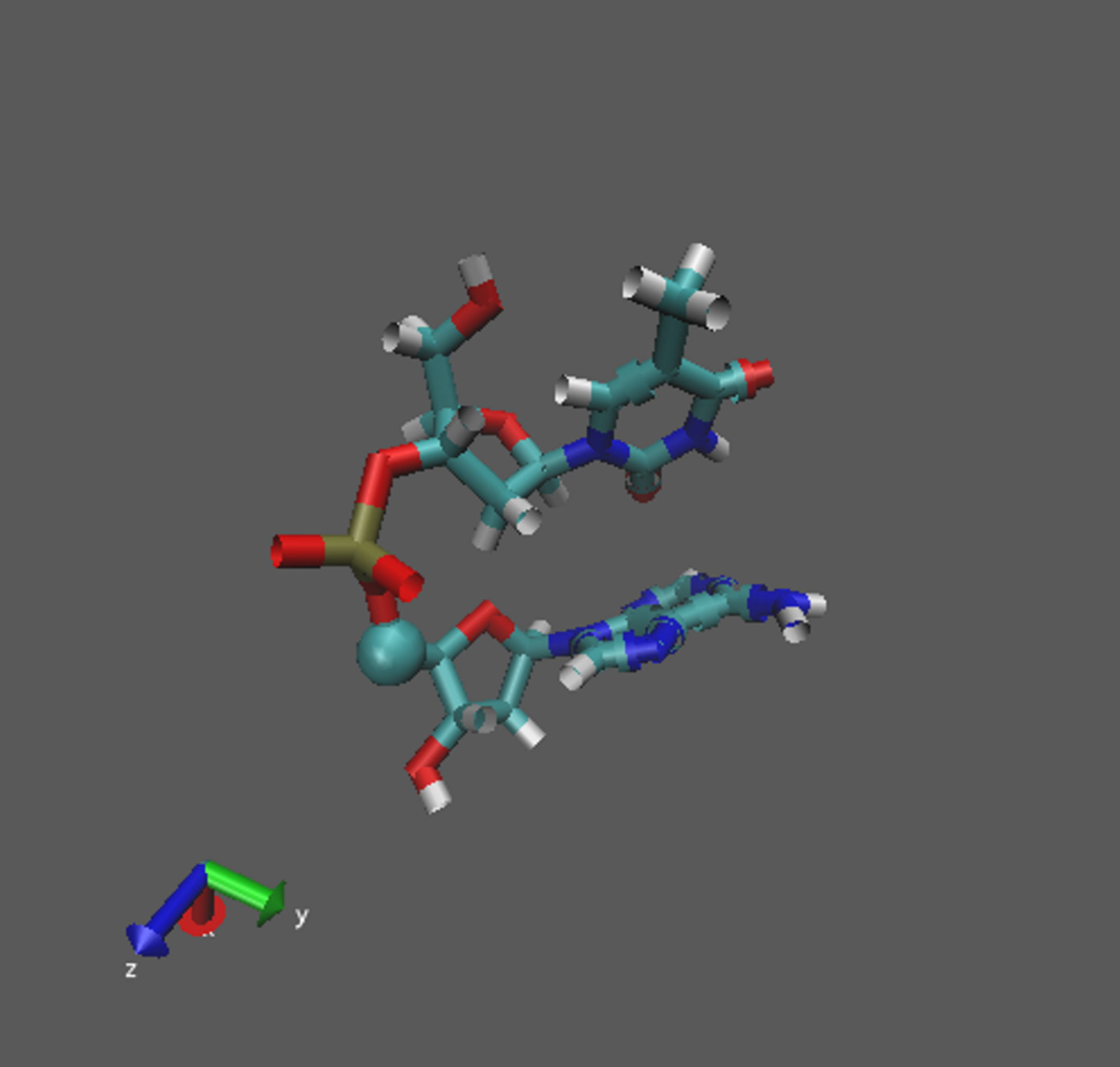}
\caption{
The example of the smallest unit of the density functional theory (DFT) simulation for $p_\textrm{L} =10$ in Fig. \ref{fig0010}.
The two hydrogens are removed from the 5$^{\prime}$ carbon between the two nucleotides A and G.
B3LYP exchange-correlation functional\cite{B3, LYP} and cc-pVDZ bases-set\cite{ccpvdz} are used in the DFL simulation with Gaussian09\cite{Gaussian}.
}
\label{fig0050}
\end{figure}

The removal of two hydrogens attached to the 5$^{\prime}$ carbon in the scar model results in a charge distribution around the removed part. 
It is, therefore, necessary to calculate the charge distribution. 
The density functional theory (DFT) calculations using Gaussian09\cite{Gaussian} are used to calculate the charge distribution.
Since DFT calculations require more computational effort than the MD simulation, calculations are performed with the smallest possible structure, taking into account the effects of the environment around the scar position. 
For example (see Fig. \ref{fig0050}), we consider the case of introducing a scar at $p_\textrm{L} =10$ in Fig. \ref{fig0010}.
Since the 5$^{\prime}$ carbon, from which two hydrogens are removed, spans nucleotides, it is reasonable to consider the two-nucleotide structure as the smallest unit to be calculated by DFT.
In the DFT simulations, we use  B3LYP exchange-correlation functional\cite{B3, LYP} and cc-pVDZ bases-set\cite{ccpvdz}.

\subsubsection{Configuration of scars in both strands}
\begin{figure}[htbp] \centering 
\includegraphics[width=12cm]{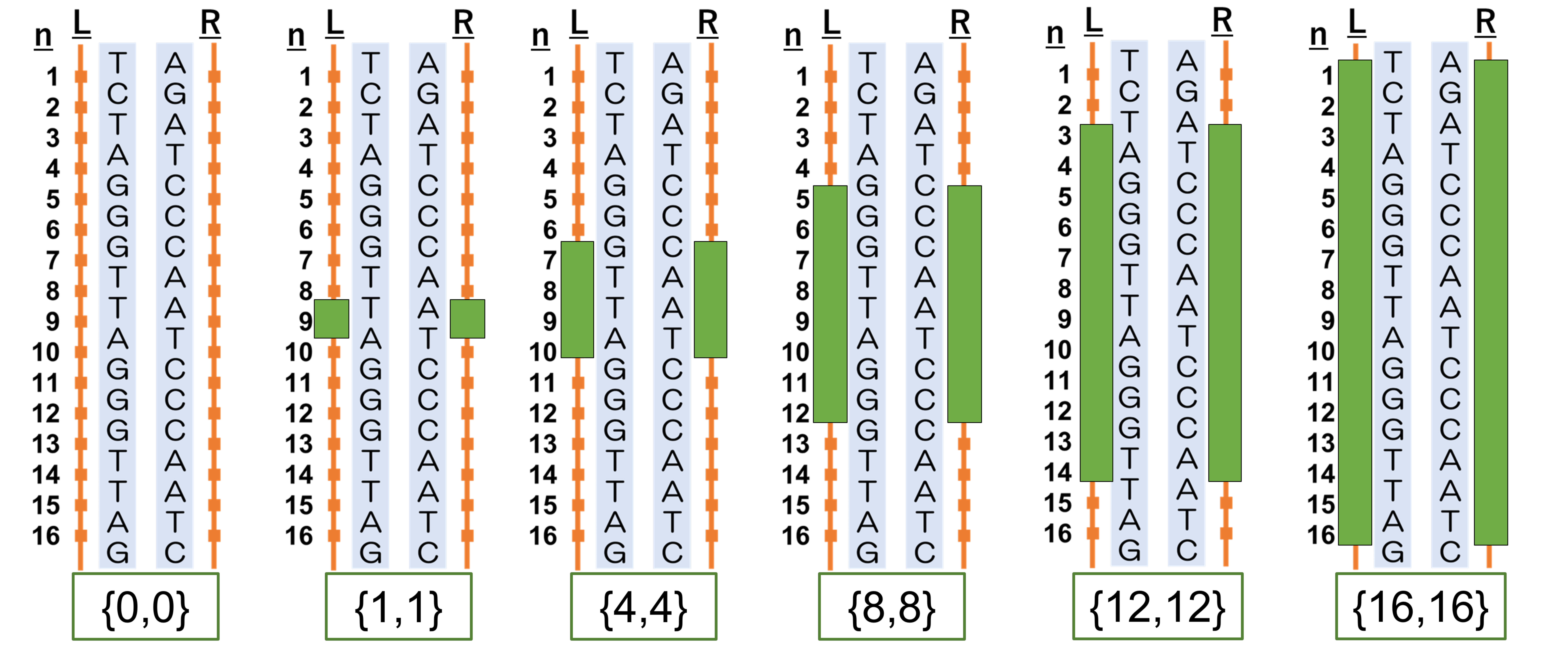}
\caption{
Schematic diagrams of the scar model with different numbers of scar pairs for $\{n_\textrm{L}, n_\textrm{R}\}$ = $\{0,0\}, \{1,1\}, $  $\{4,4\}, \{8,8\}, \{12,12\}, $ and  $\{16,16\}$.
Each green squares represent scars.
}
\label{fig0060}
\end{figure}

\begin{figure}[htbp] \centering 
\includegraphics[width=12cm]{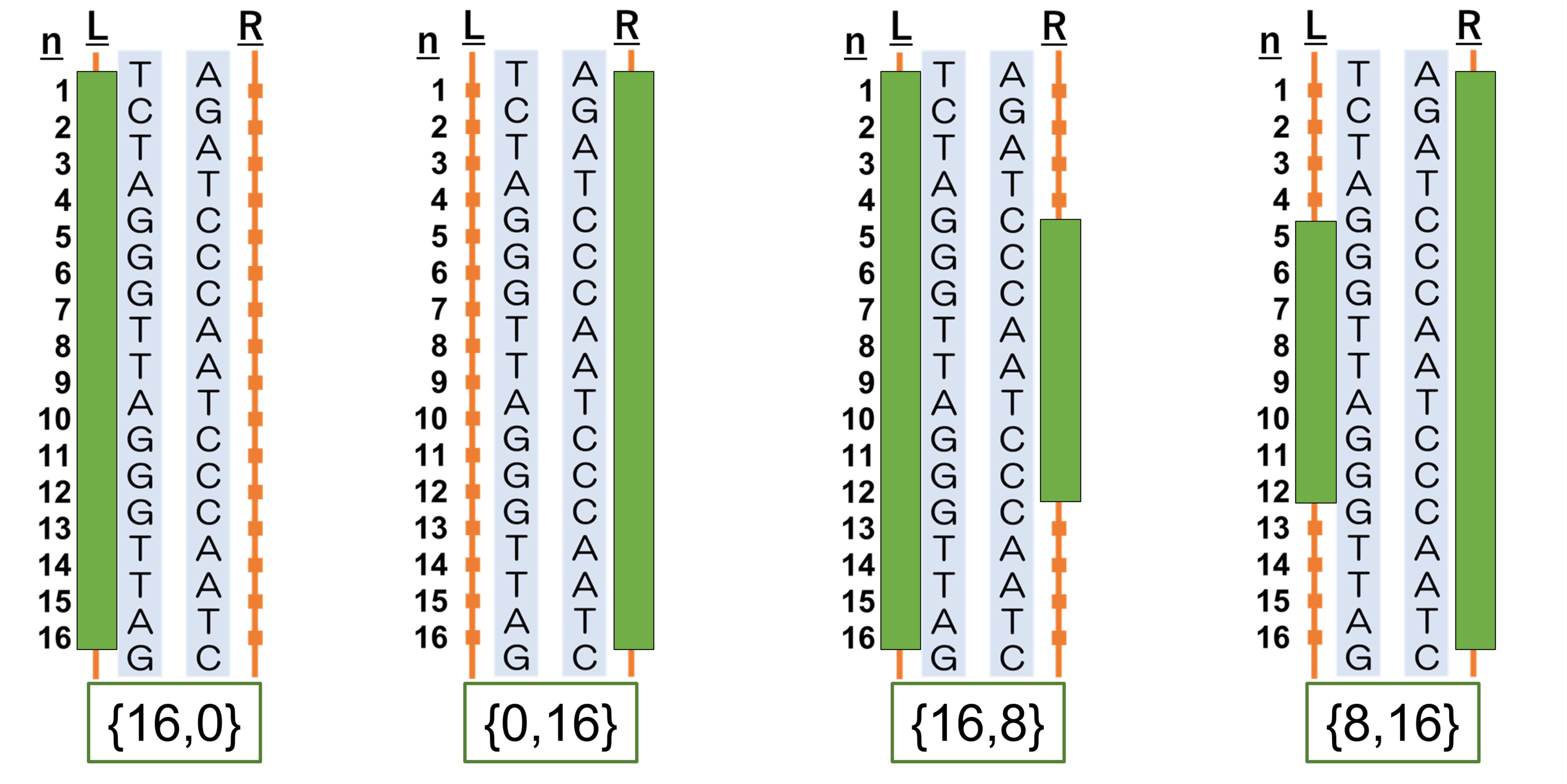}
\caption{
Schematic diagrams of the scar model with different numbers of scar pairs for $\{n_\textrm{L}, n_\textrm{R}\}$ = $\{16,0\}, \{0,16\}, \{16,8 \}$ and $\{8,16\}$.
Each green squares represent scars, as in Fig. \ref{fig0060}.
}
\label{fig0070}
\end{figure}

Ten patterns of simulations were performed to investigate the effect of scars on the telomeric DNA, varying the number and location of scars. 
To distinguish between them, the number of scratches in the left or right strands is defined as $n_\textrm{L}$ or $n_\textrm{R},$ respectively. 
We consider the arrangements $\{n_\textrm{L}, n_\textrm{R}\}$ = $\{0,0\}, \{1,1\}, $ $\{4,4\}, \{8,8\}, \{12,12\}, $ $\{16,16\}, \{16,0\}, \{0,16\}, \{16,8 \}$ and $\{8,16\}$ as shown in Figs. \ref{fig0060} and \ref{fig0070}.

\section{Molecular Dynamics Simulation Results}
\begin{figure}[htbp] \centering 
\includegraphics[width=6cm]{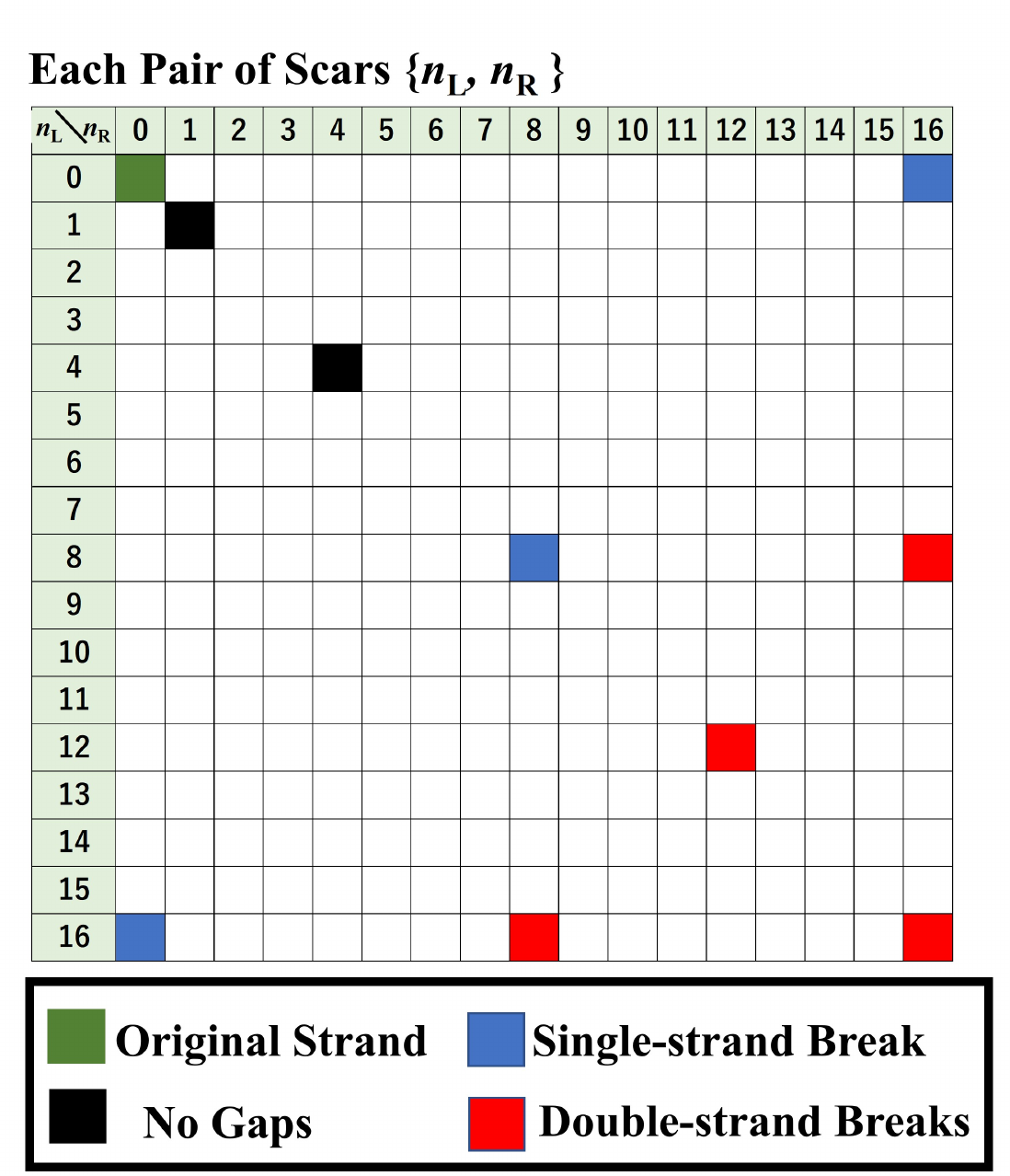} 
\caption{
Summary of the MD simulation results for $\{n_\textrm{L}, n_\textrm{R} \}$ = $\{0,0\}, \{1,1\}, $ $\{4,4\}, \{8,8\}, \{12,12\}, $ $\{16,16\}, \{16,0\}, \{0,16\}, \{16,8 \}$ and $\{8,16\}$ as shown in Figs. \ref{fig0060} and \ref{fig0070}.
The green square represents telomeric DNA in the absence of scars. 
At the end of the simulation ($t = $ 11.4 ns), the black squares indicate no gaps in both DNA strands, the blue ones indicate single-strand breaks (SSB), and the red ones indicate double-strand breaks (DSBs).
}
\label{fig0080}
\end{figure}

MD simulations were performed using the 10 different scar models proposed in the previous section  (Figs. \ref{fig0060} and \ref{fig0070})  to investigate their dependence on the number and location of scars.
As the strand break is transient and recombines rapidly, the strand break longer than 0.01 ns is defined as a `gap,' the gap occurring on only one strand as single-strand breaks (SSB), and the gaps occurring on both strands are defined as double-strand breaks (DSBs) and the results are shown in Fig. \ref{fig0080}.

\begin{figure}[htbp] \centering
\includegraphics[width=6cm]{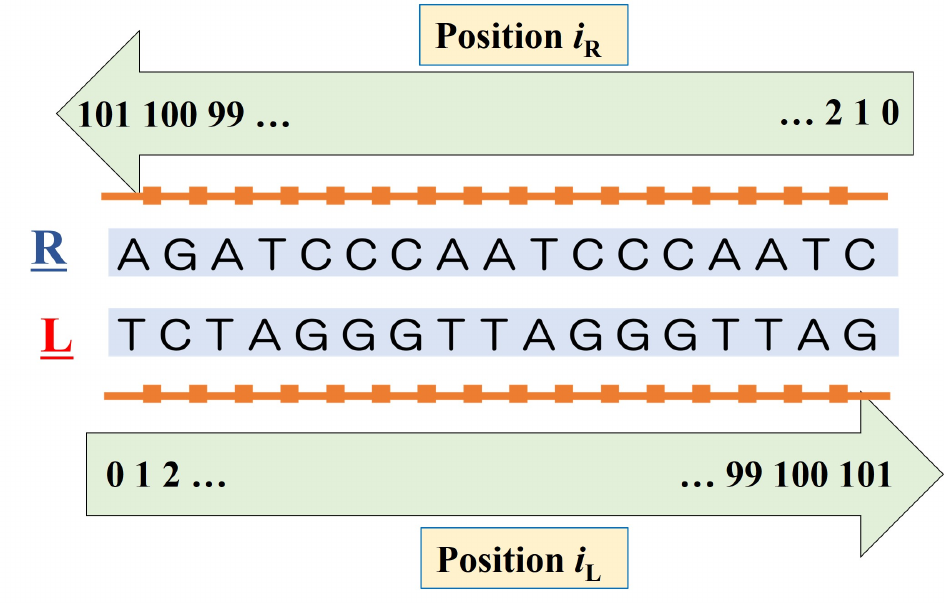}
\caption{
Definition of the position $i_\textrm{R, L}$ on each R or L strand.
The $i_\textrm{R, L}$ means atoms in the backbone of the R or L strand. 
On the other hand,  the position $p_\textrm{R, L}$ in Fig. \ref{fig0010} means the positions of only phosphorous atoms in the backbone of the R or L strand.
}
\label{fig0090}
\end{figure}

Until  $t = 11.4$ ns, DSBs occur in the four cases $\left\{12, 12\right\}, \left\{16, 16\right\},$ $\left\{16, 8\right\}$ and $\left\{8, 16\right\},$ and SSB occurs in the three cases $\left\{8, 8\right\},$ $\left\{16, 0\right\}$ and $\left\{0, 16\right\}.$
In particular, for $\left\{16, 0\right\},$ and $\left\{0, 16\right\},$ gaps occur only on the scarred side of the chain. 
No gaps occur within the simulation time in $\left\{1, 1\right\}$ and $\left\{4, 4\right\}.$
Obviously, in the no scarred `original' telomeric DNA case, \textit{i. e.}, $\left\{0, 0\right\},$ no gaps appear in the DNA, and the double-strand structure is preserved.  
The above results can be summarised by the total number of scars  $n_\textrm{L} +  n_\textrm{R}$ as follows:
\begin{enumerate}
\item{$n_\textrm{L} +  n_\textrm{R}$ = 2 and 8: No gaps appear.}
\item{$n_\textrm{L} +  n_\textrm{R}$ = 16: Single-strand break (SSB) occurs.}
\item{$n_\textrm{L} +  n_\textrm{R}$ = 24 and 32: Double-strand breaks (DSBs) occur.}
\end{enumerate}
Thus, it is possible to show that the total number of scars $n_\textrm{L} +  n_\textrm{R}$ can be a good indicator related to the telomeric DNA strand breaks.
We also summarise, in Table \ref{tab:result}, the time of occurrence of SSB and DSBs,  and the total number of gaps generated in the final state ($t=$ 11.4 ns) for each $\left\{ n_\textrm{L},   n_\textrm{R} \right\}$ pair.
In the subsequent Subsections \ref{original}, \ref{dsbs}, \ref{ssb}, and \ref{nogap}, the following four physical quantities are presented and discussed for each $\left\{ n_\textrm{L},   n_\textrm{R} \right\}$ pair.
\begin{description}
\item[(i)] The final molecular structure at $t=$ 11.4 ns. 
\item[(ii)] The location of $p_\textrm{R, L} $ in the telomeric DNA where gaps appear at $t=$ 11.4 ns.
\item[(iii)] The time evolution of the gap number from 10.7 ns to 11.4 ns.
\item[(iv)] The spatial ($i_\textrm{R, L}$) distributions of the bond order at $t=$ 10.7 ns and 11.4 ns. 
Here, the spatial position $i_\textrm{R, L}$ is defined in Fig. \ref{fig0090}.
Moreover, bond order is a formal measure of the multiplicity of covalent bonds between two atoms.
\end{description}
Here we define the bond order, which is calculated directly from an interatomic distance using the empirical formula:
\begin{eqnarray}
  \mathrm{BO}_{ ij} &=& \mathrm{BO}_{ ij}^{\sigma} + \mathrm{BO}_{ ij}^{\pi}\ + \mathrm{BO}_{ ij}^{\pi\pi}\nonumber  \\
  &=& \exp\left[p_1 \cdot \left( \frac{r_{ij}}{r^{\sigma}_{0}}\right)^{p_2}\right]
   + \exp\left[ p_3 \cdot \left(\frac{r_{ij}}{r^{\pi}_{0}}\right)^{p_{4}}\right]
   + \exp \left[ p_5 \cdot \left( \frac{r_{ij}}{r^{\pi \pi}_{0}}\right)^{p_6} \right] , \label{BO}
\end{eqnarray}
where BO is the bond order between atoms $i$ and $j$, $r_{ij}$ is interatomic distance, $r^{\sigma, \pi, \pi\pi }_0 $ terms are equilibrium bond lengths, and $p_{1, 2,  \cdots, 6 }$ terms are empirical parameters\cite{bond}.

\begin{table}[htbp] 
  \begin{center}
  \caption{Times of single-strand break (SSB) and double-strand breaks (DSBs) occurrence, and the number of gaps at $t=11.4$ ns. }
  \begin{tabular}{|c|c|c||>{\centering}p{4em}|>{\centering}p{4em}|c|}\hline
 \multicolumn{3}{|c||}{Scars in Backbone} &  \multicolumn{2}{c|}{Occurrence Time} & \multicolumn{1}{c|}{Num. of Gaps}   \\ \cline{1-3} 
Case & Pair  & Total Num.  &  \multicolumn{2}{c|}{[ns] } & \multicolumn{1}{c|}{at}   \\ \cline{4-5} 
 & $\left\{n_\textrm{L},n_\textrm{R}\right\}$ &  $n_\textrm{L}+n_\textrm{R}$  &   \multicolumn{1}{c|}{SSB}  & \multicolumn{1}{c|}{DSBs} & \multicolumn{1}{c|}{$t=$ 11.4 ns} \\ \hline \hline
  O  & $\rm{\left\{0,0\right\}}$ & 0 & 11.357 & null & 2\\ \Hline
  D1 & $\rm{\{16,16\}}$&32 & 10.774 & 11.242 & 8\\ \hline
  D2 & $\rm{\{16,8\}}$ &24& 10.904 & 11.105 & 6\\ \hline
  D3 & $\rm{\{8,16\}}$ &24& 11.039 & 11.041 & 7\\ \hline
  D4 & $\rm{\{12,12\}}$&24 & 10.884 & 11.160 & 4\\ \Hline
  S1 & $\rm{\{16,0\}}$ &16& 10.904 & null & 9\\ \hline
  S2 & $\rm{\{0,16\}}$ &16& 10.993 & null & 3\\ \hline  
  S3 & $\rm{\{8,8\}}$ &16& 10.945 & null & 1\\ \Hline
  N1 & $\rm{\{4,4\}}$ &8& null &null & null \\ \hline  
  N2 & $\rm{\{1,1\}}$ &2& null & null & null \\ \hline
  \end{tabular}
  \label{tab:result}
  \end{center}
\end{table}

\subsection{Original Strand: O case}\label{original}
First, the time evolution of the telomeric DNS structure is reported for no scars cases, \textit{i. e.},  $n_\textrm{L}+n_\textrm{R}$ = 0 (O case in Table. \ref{tab:result}).
This behavior is the basis for discussing the results in the scarred cases.

\begin{figure}[htbp] 
\includegraphics[width=9cm]{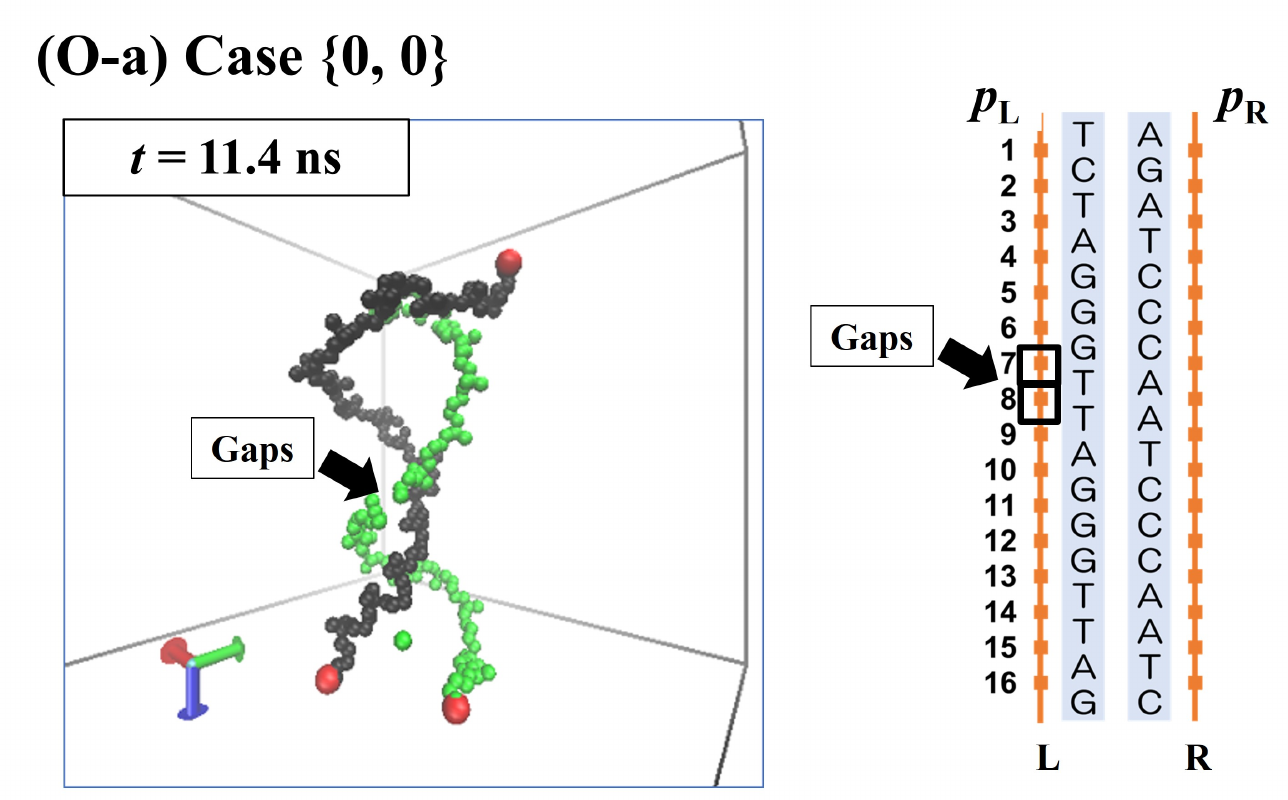} \\
\includegraphics[width=14cm]{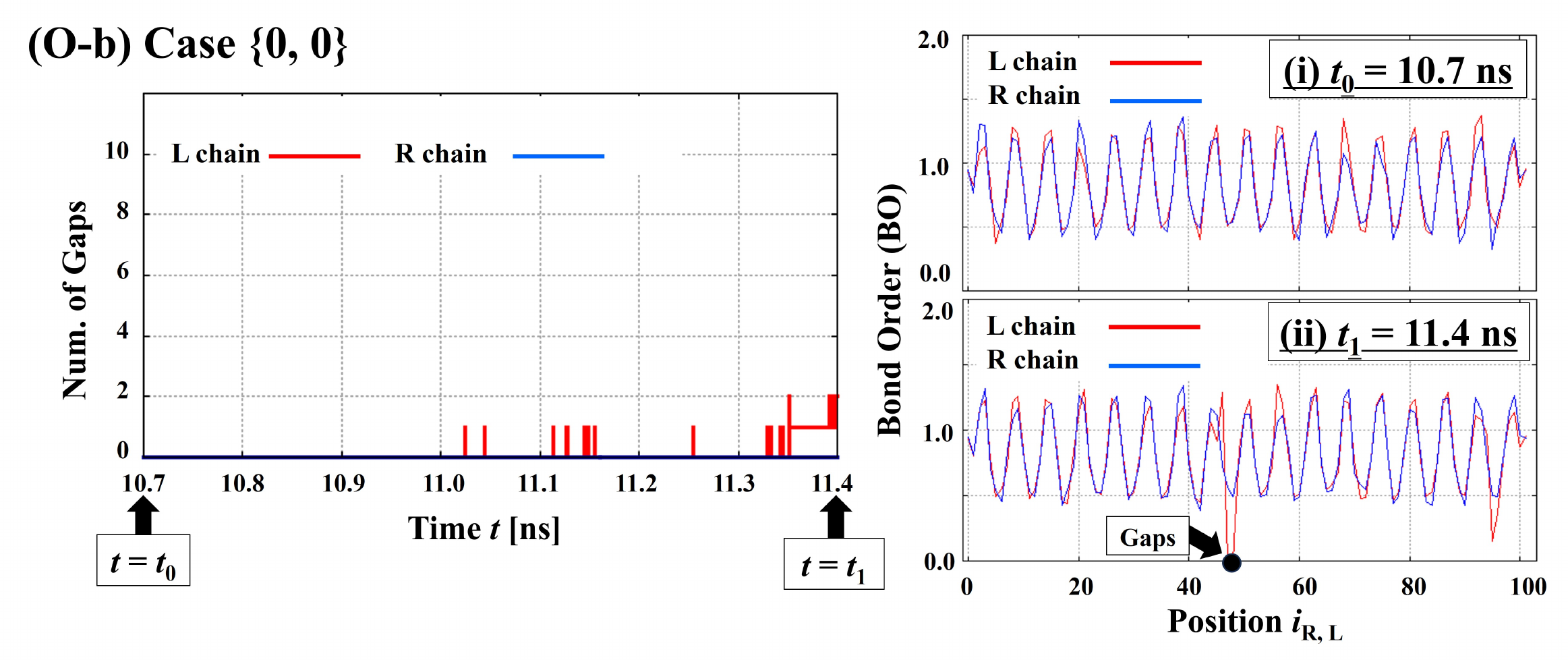}\caption{
For O case $\left\{0, 0\right\},$ the final ($t=$ 11.4 ns) molecular structure is depicted on the left side of (O-a).
There, one gap appears in the L-strand (green chain), the position of which is represented by black squares in the right figure of (O-a) with $p_\textrm{L} = $ 7 and 8.
The left side of (O-b) shows the time evolution of the number of gaps in each L (red) or R (blue) strand from $t_0 = $10.7 ns to $t_1 = $11.4 ns. 
The spatial distribution of the bond order for each strand is drawn on the right side of (O-b) at $t = t_0$ (upside) and $t_1$ (downside).
Spatial distribution here means distribution with respect to $i_\textrm{R, L},$ which means the position of atoms in the backbone of the R or L strand as shown in Fig. \ref{fig0090}. 
This figure shows that there is a gap (black circle) in the L strand at $t_1$ (downside).
}
\label{resO}
\end{figure}

From Fig. \ref{resO}, it is shown that one gap appears in the L-strand (green chain), the position of which is represented by black squares in the right figure of (O-a) with $p_\textrm{L} = $ 7 and 8.
In addition, the time evolution of the number of gaps in the left-hand side of Fig. \ref{resO} shows how the gaps are repeatedly created and extinguished until finally there are only one or two gaps, which eventually remain. 
This time evolution shows that thermal fluctuations can cause gaps in our telomeric  DNA by the simulation for the 0.7 ns period from $t_0 =$ 10.7 ns to $t_1 = $ 11.4 ns, even if the telomeres were not initially scarred.
However, it should also be noted that this gap may result in a single-strand break but not double-strand breaks.
This must be taken into account when analyzing other cases of scars.

\subsection{Double Strand Breaks (DSBs): D1, D2, D3 and D4 cases}\label{dsbs}
\begin{figure}[htbp] 
\includegraphics[width=9cm]{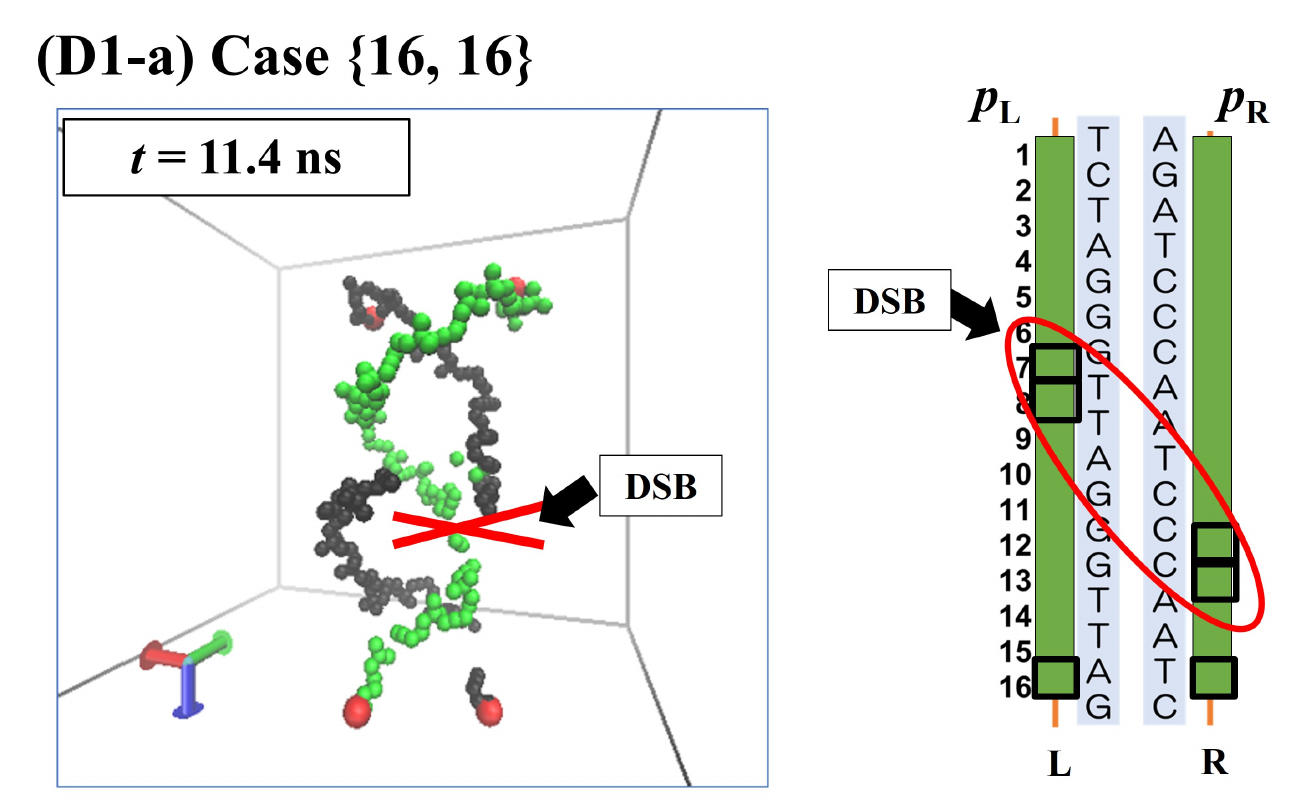} \\
\includegraphics[width=14cm]{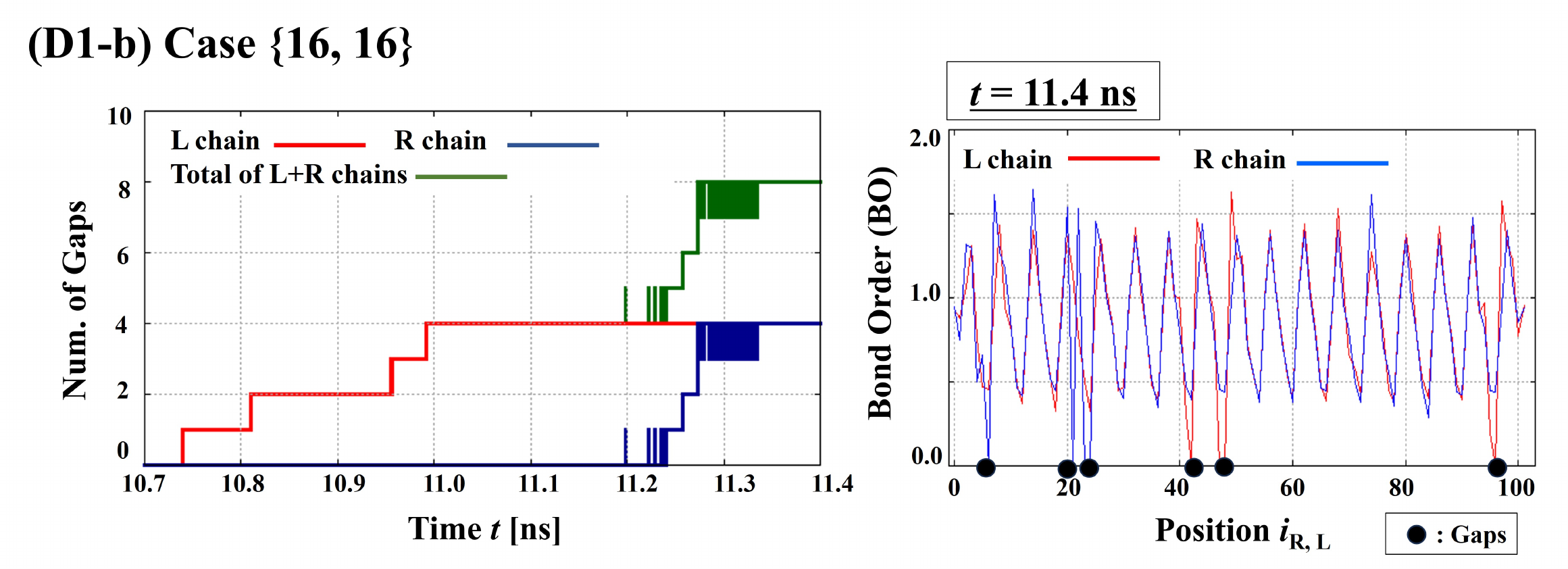} 
\caption{
In D1 case $\left\{16, 16\right\},$ there are 16 scars each on the R and L chains (32 scars in total), represented by green squares on the right-hand side of (D1-a).
The red oval means the double-strand breaks.
Moreover, double-strand breaks, which are illustrated by the red line, appear in the final molecular structure on the left-hand side of (D1-a).
The figure (D1-b) shows the time evolution of the number of gaps and the spacial distribution of the bond order. 
}
\label{resD1}
\end{figure}

\begin{figure}[htbp] 
\includegraphics[width=9cm]{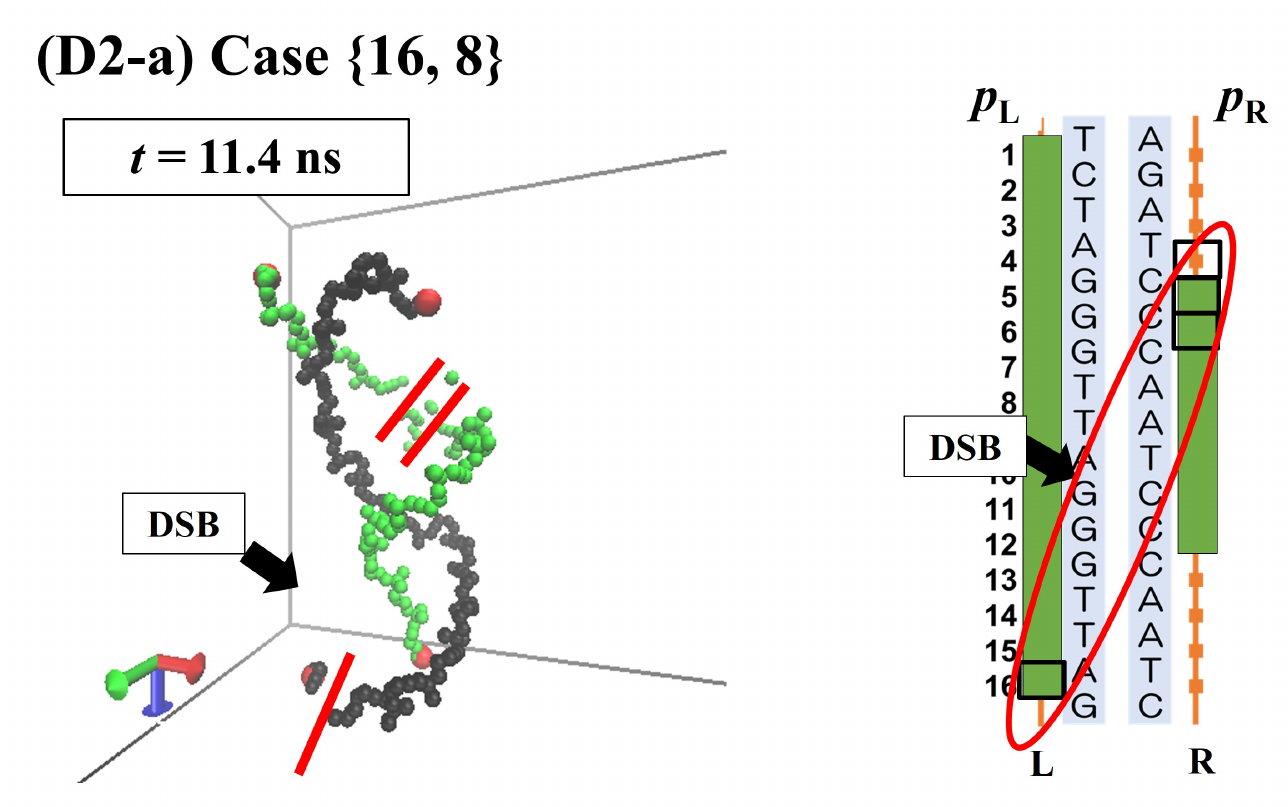} \\ 
\includegraphics[width=14cm]{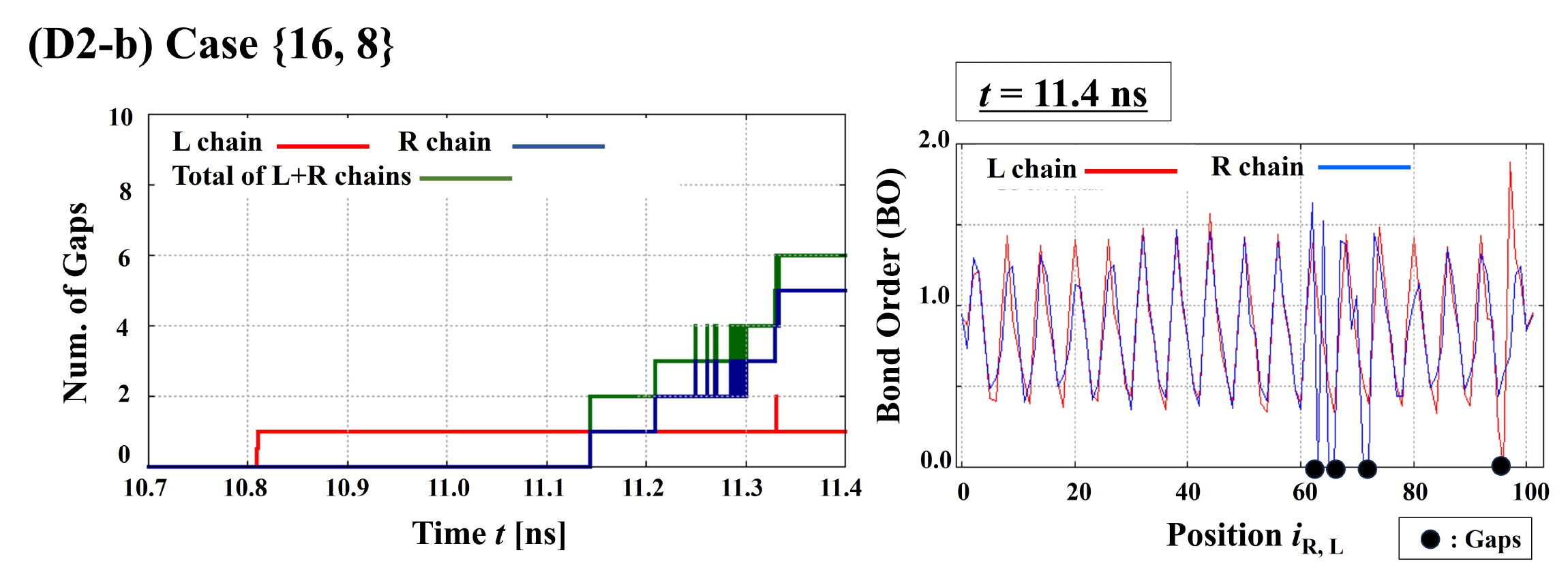} 
\caption{
In D2 case $\left\{16, 8\right\}$, there are 16 scars on the R chain and 8 scars on the L chain (24 scars in total), represented by green squares on the right-hand side of (D2-a).
The red oval means the double-strand breaks.
Moreover, double-strand breaks, which are illustrated by the red line, appear in the final molecular structure on the left-hand side of (D2-a).
The figure (D2-b) shows the time evolution of the number of gaps and the spacial distribution of the bond order. 
}
\label{resD2}
\end{figure}

\begin{figure}[htbp] 
\includegraphics[width=9cm]{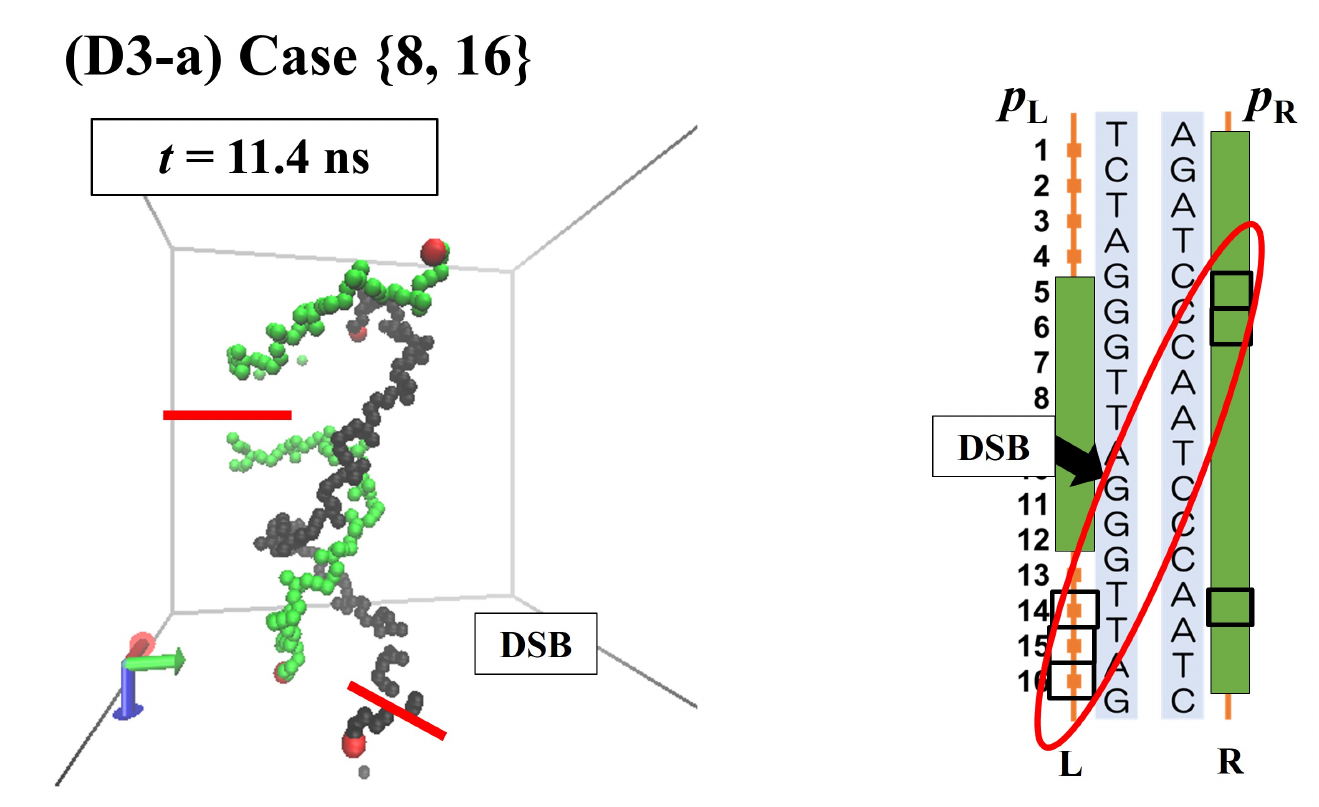} \\
\includegraphics[width=14cm]{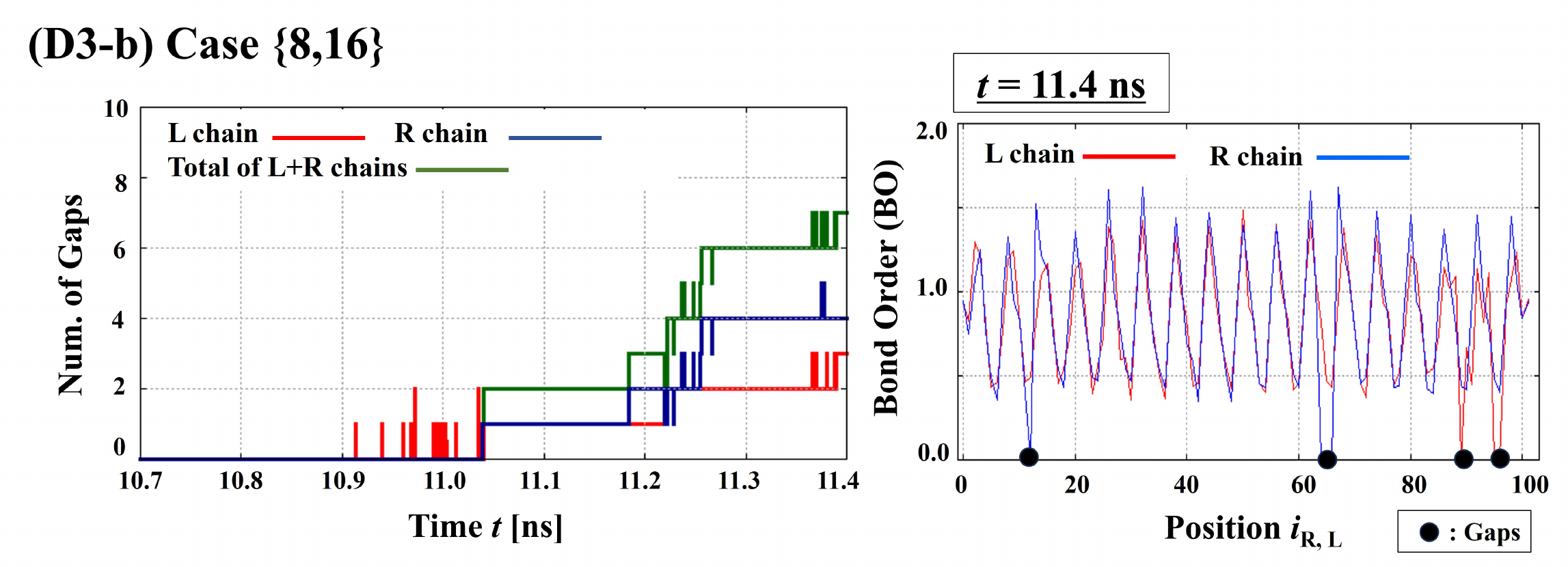} 
\caption{
In D3 case $\{8, 16\},$ there are 8 scars on the R chain and 16 scars on the L chain (24 scars in total), represented by green squares on the right-hand side of (D3-a).
In this case, the left and right sides of case D2 are reversed.
Moreover, double-strand breaks, which are illustrated by the red line, appear in the final molecular structure on the left-hand side of (D3-a).
The figure (D3-b) shows the time evolution of the number of gaps and the spacial distribution of the bond order. 
}
\label{resD3}
\end{figure}

\begin{figure}[htbp] 
\includegraphics[width=9cm]{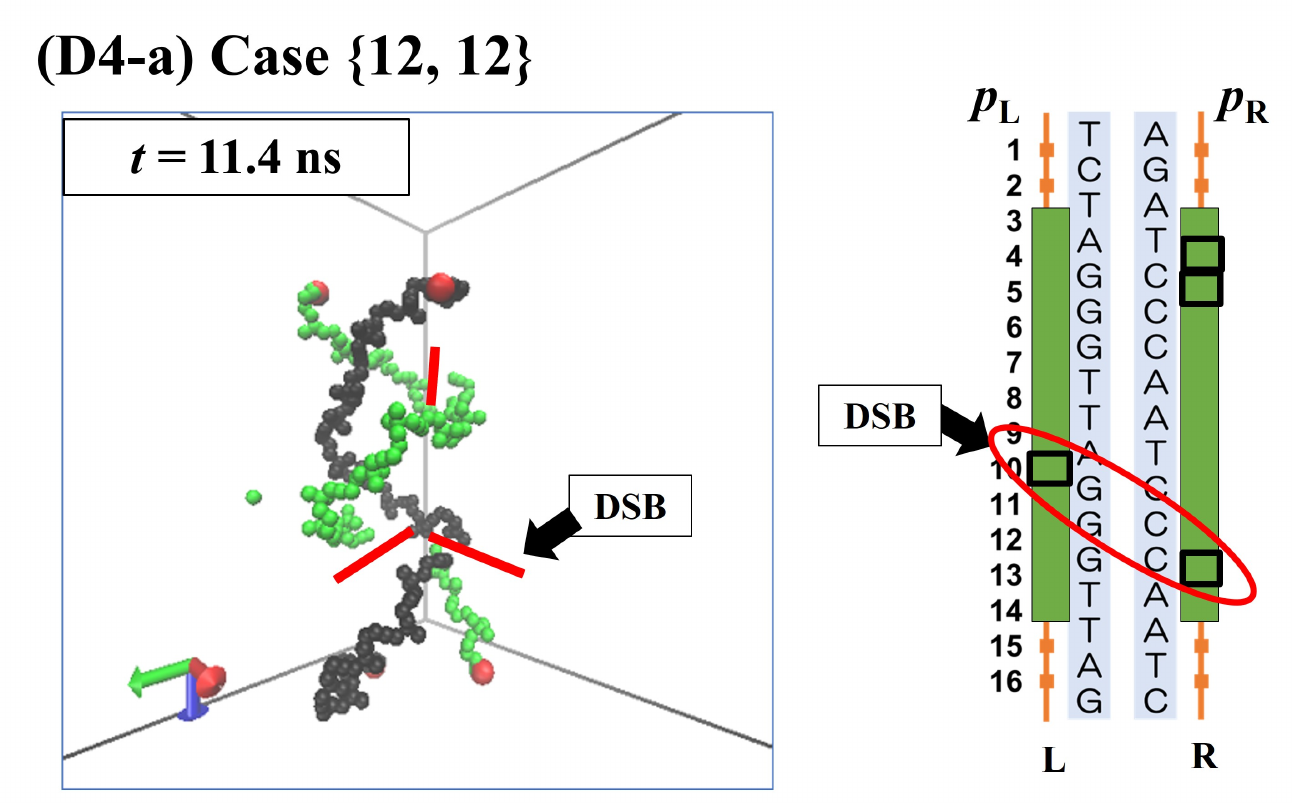} \\
\includegraphics[width=14cm]{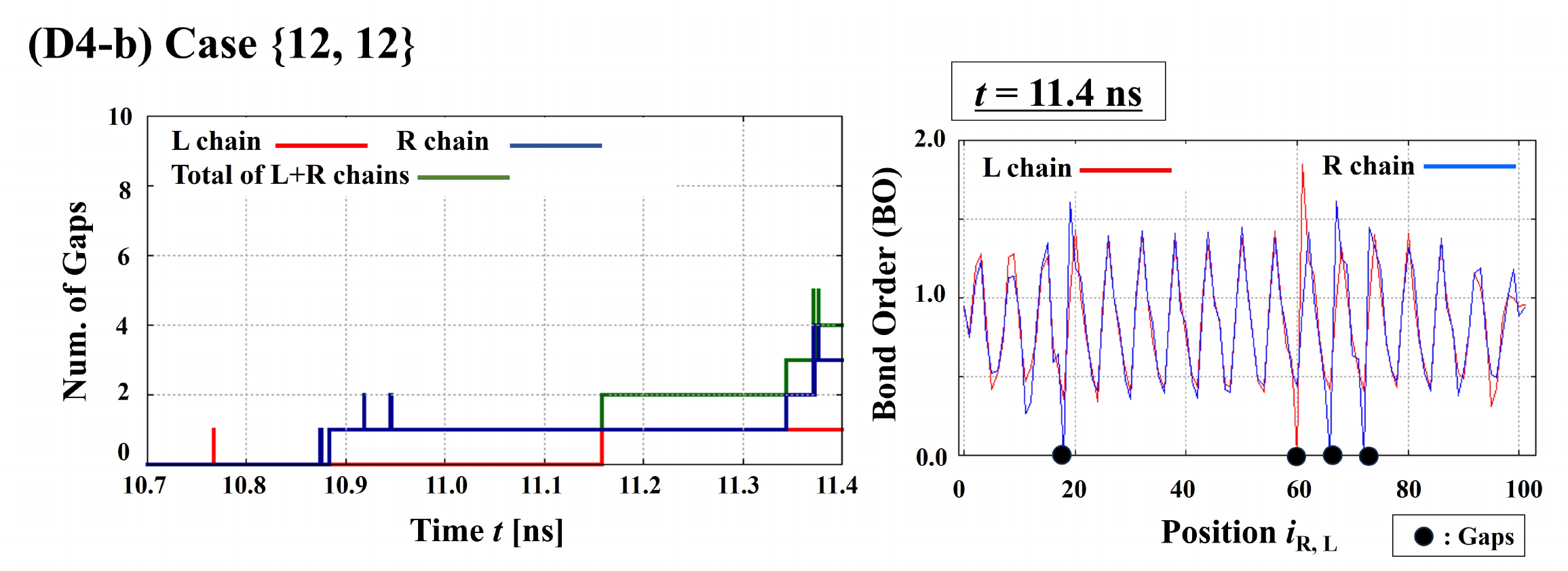}
\caption{
In D4 case $\{12, 12\}, $  there are 12 scars each on the R and L chains (24 scars in total).
Although the total number of scars is less than in other cases D1, D2, and D3, DSBs are still occurring.
}
\label{resD4}
\end{figure}

In these cases, \textit{i. e.,} $n_\textrm{L} +  n_\textrm{R}$ = 24 and 32, the double-strand breaks occur as shown in Figs. \ref{resD1}, \ref{resD2}, \ref{resD3} and \ref{resD4}. 
The observation that once a gap is made it does not heal, as obtained from all the time evolution of the number of scars on the strands in Figs. \ref{resD1}(D1-b),  \ref{resD2}(D2-b), \ref{resD3}(D3-b) and \ref{resD4}(D4-b), shows that the gaps are expanding. 
This phenomenon occurs in both strands, resulting in the DSBs. 
This behavior is very different from the repetition of creation and recovery of gaps due to thermal fluctuations in the O case in the section \ref{original}, where the gaps do not grow.

\subsection{Single Strand Breaks (SSB): S1, S2 and S3 cases}\label{ssb}
\begin{figure}[htbp] 
\includegraphics[width=9cm]{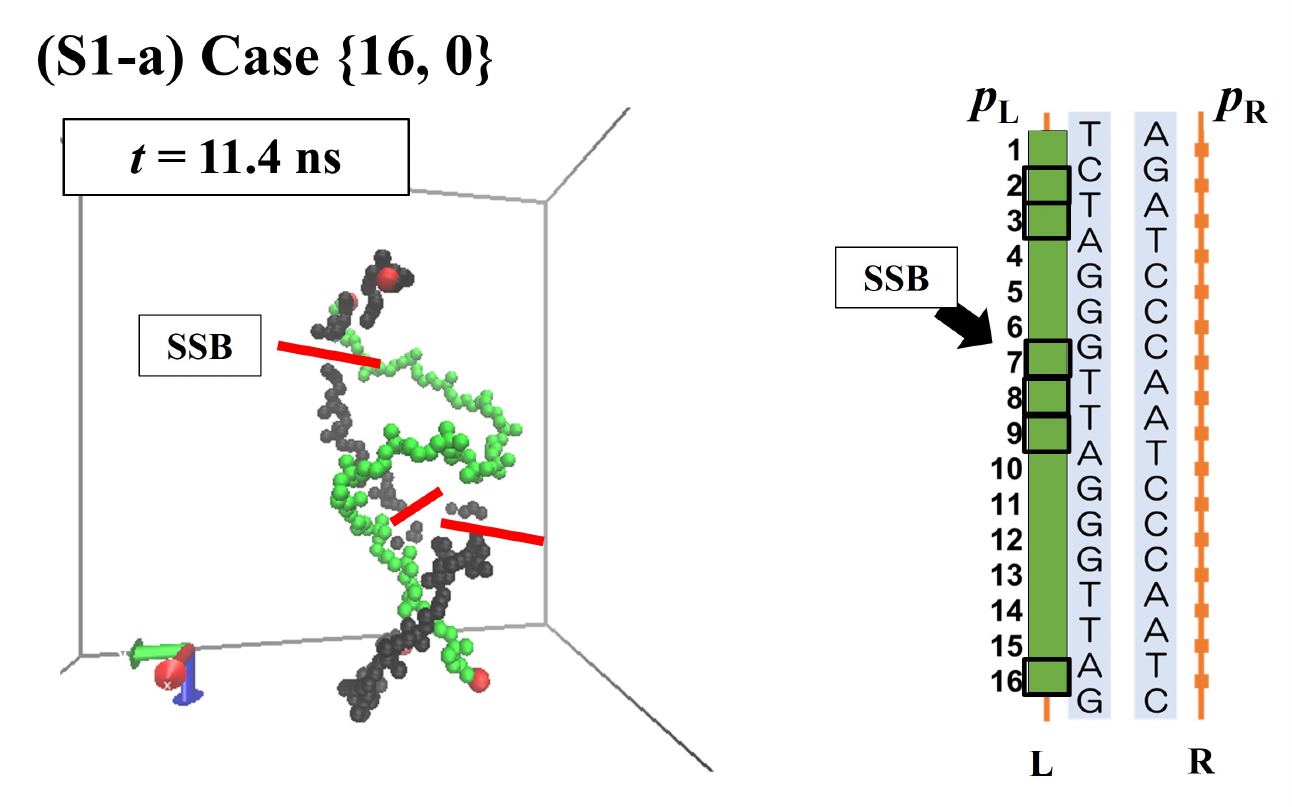} \\
\includegraphics[width=14cm]{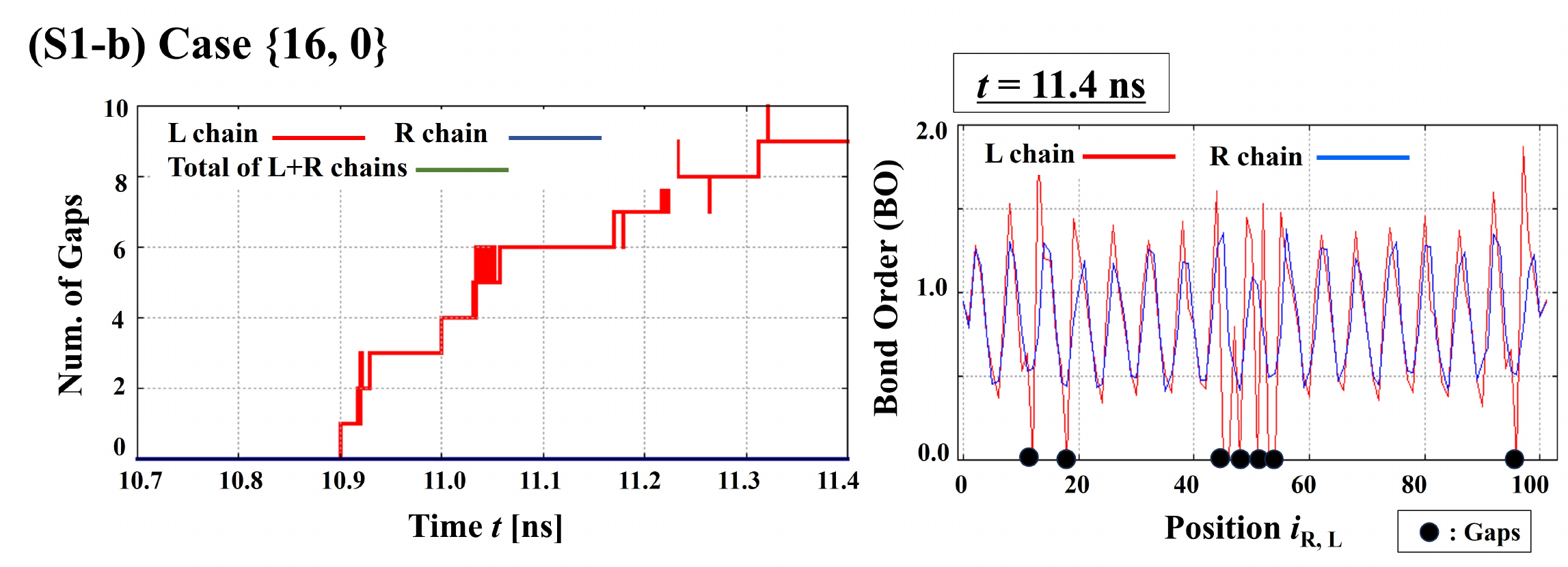} 
\caption{
In S1 case $\{16, 0\}, $ there are 16 scars on the L chain and no scars on the R chain (16 scars in total), represented by green squares on the left-hand side of (S1-a).
The gaps in the L chain increase as time goes by. 
In the R chain, on the other hand, no gaps occur.
}
\label{resS1}
\end{figure}

\begin{figure}[htbp] 
\includegraphics[width=9cm]{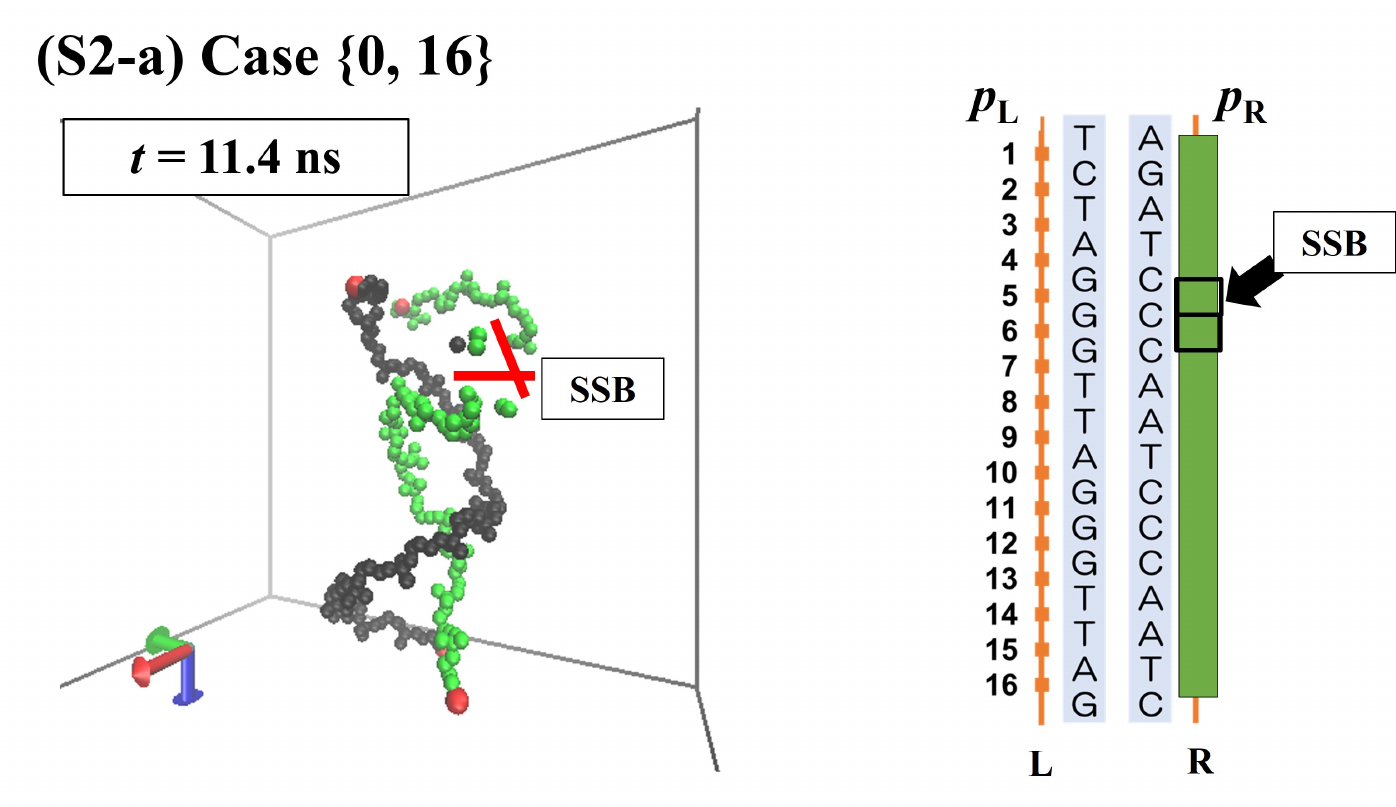} \\
\includegraphics[width=14cm]{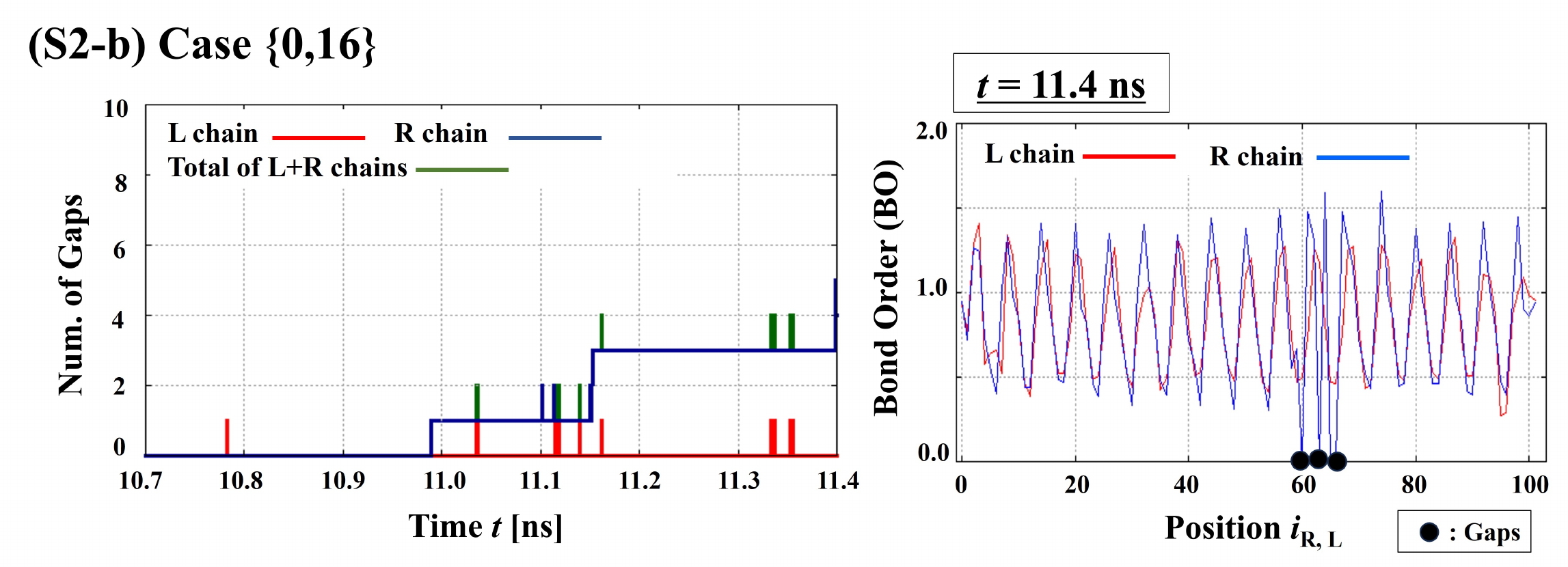} 
\caption{
In S2 case $\{0, 16\}, $ there are 16 scars on the R chain and no scars on the L chain (16 scars in total) which is contrary to the S3 case, represented by green squares on the right-hand side of (S2-a).
The gaps in the R chain increase as time goes by. 
In the L chain, on the other hand, gaps occur momentarily but are quickly recovered.
}
\label{resS2}
\end{figure}

\begin{figure}[htbp] 
\includegraphics[width=9cm]{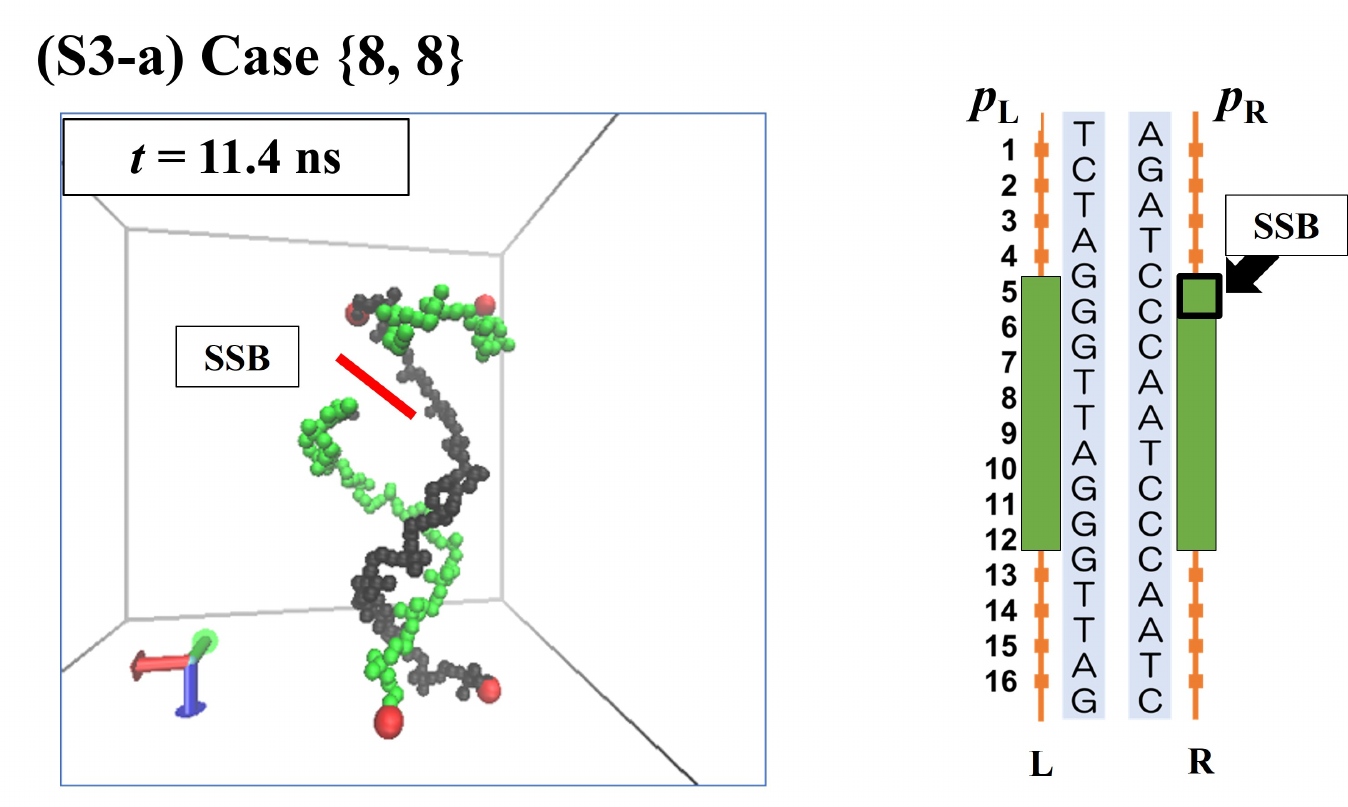} \\
\includegraphics[width=14cm]{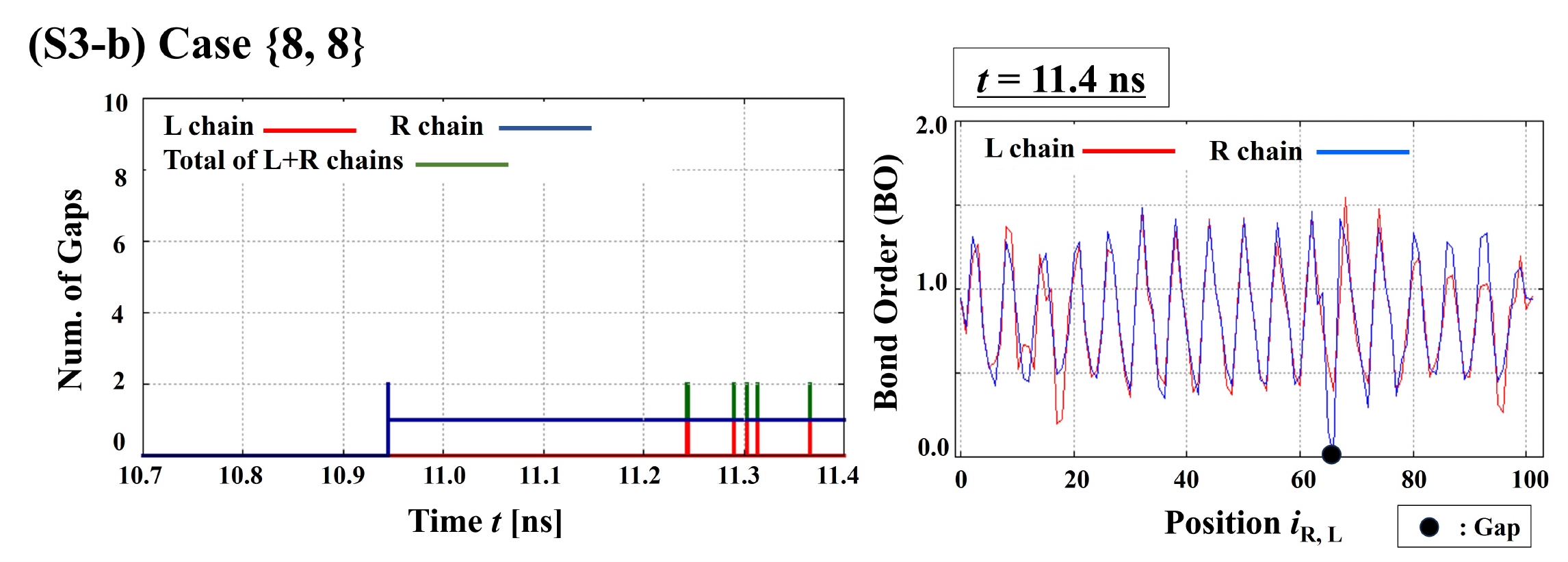} 
\caption{
In S3 case $\{8, 8\}, $ there are 8 scars each on the R and L chains (16 scars in total).
Gaps occur in the R chain, with one failing to recover and persisting. 
In contrast, once a gap emerges in the L chain, it is capable of recovery. 
This phenomenon results in the formation of an SSB.
}
\label{resS3}
\end{figure}

In these cases, \textit{i. e.,} $n_\textrm{L} +  n_\textrm{R}$ = 16, the single-strand break occurs as shown in Figs. \ref{resS1}, \ref{resS2} and \ref{resS3}. 
Comparing the three cases, S1, S2 and S3, the number of gaps becomes larger in the order S3 $<$ S2 $<$ S1.
In all cases, only one single-strand break occurs.

\subsection{No Gaps: N1 and N2 cases}\label{nogap}
\begin{figure}[htbp] 
\includegraphics[width=9cm]{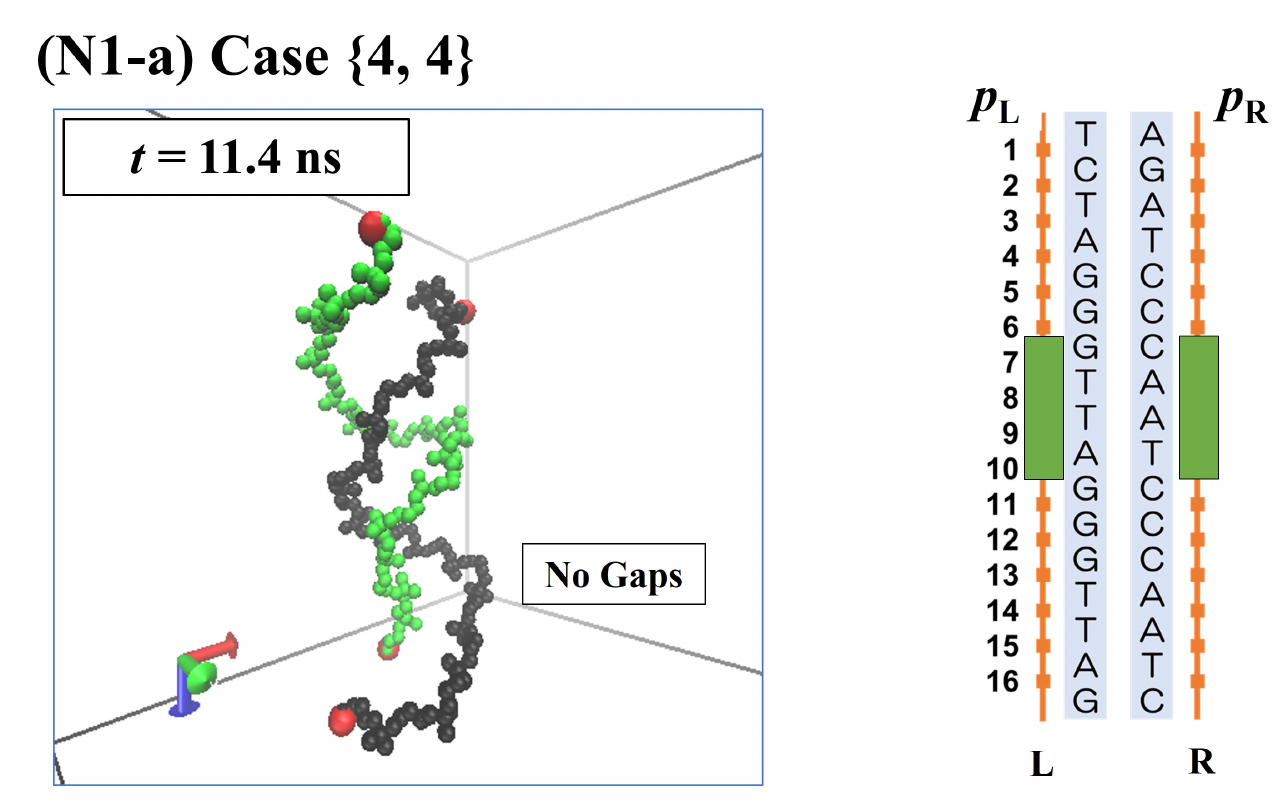} \\
\includegraphics[width=14cm]{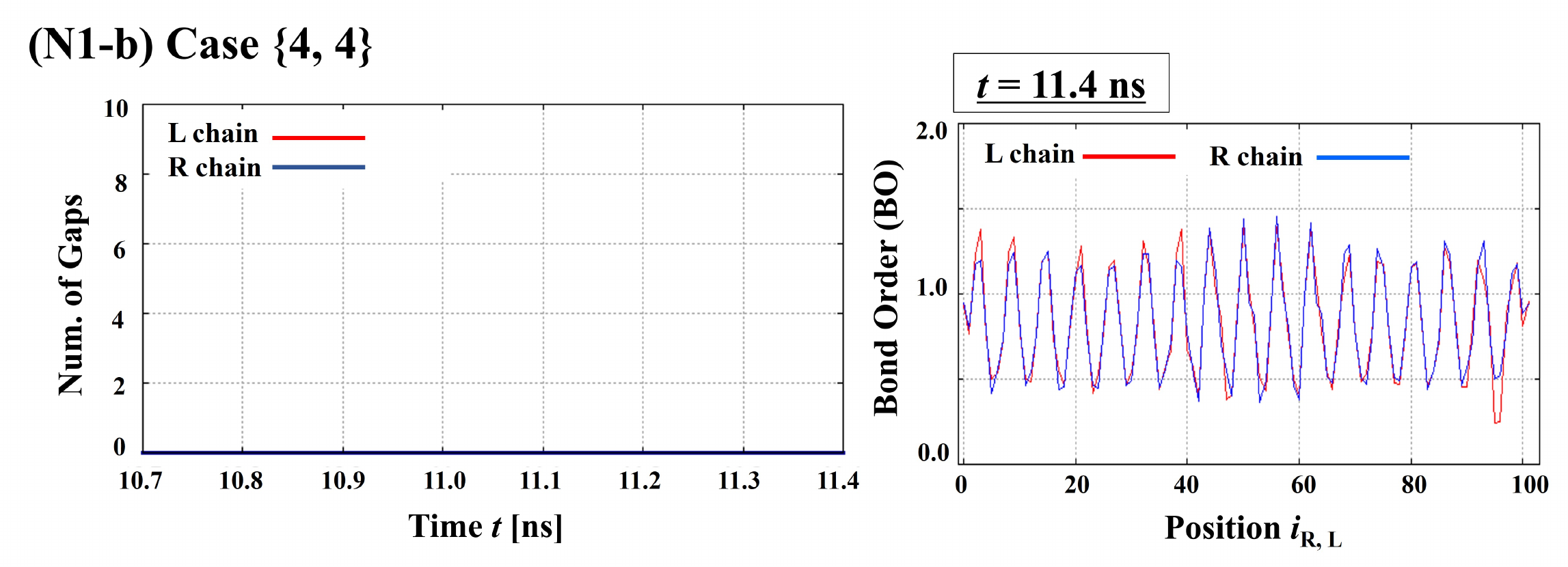}
\caption{
In N1 case $\{4, 4\},$ there are 4 scars each on the R and L chains (8 scars in total).
No gaps occur in the R and L chains. 
}
\label{resN1}
\end{figure}

\begin{figure}[htbp] 
\includegraphics[width=9cm]{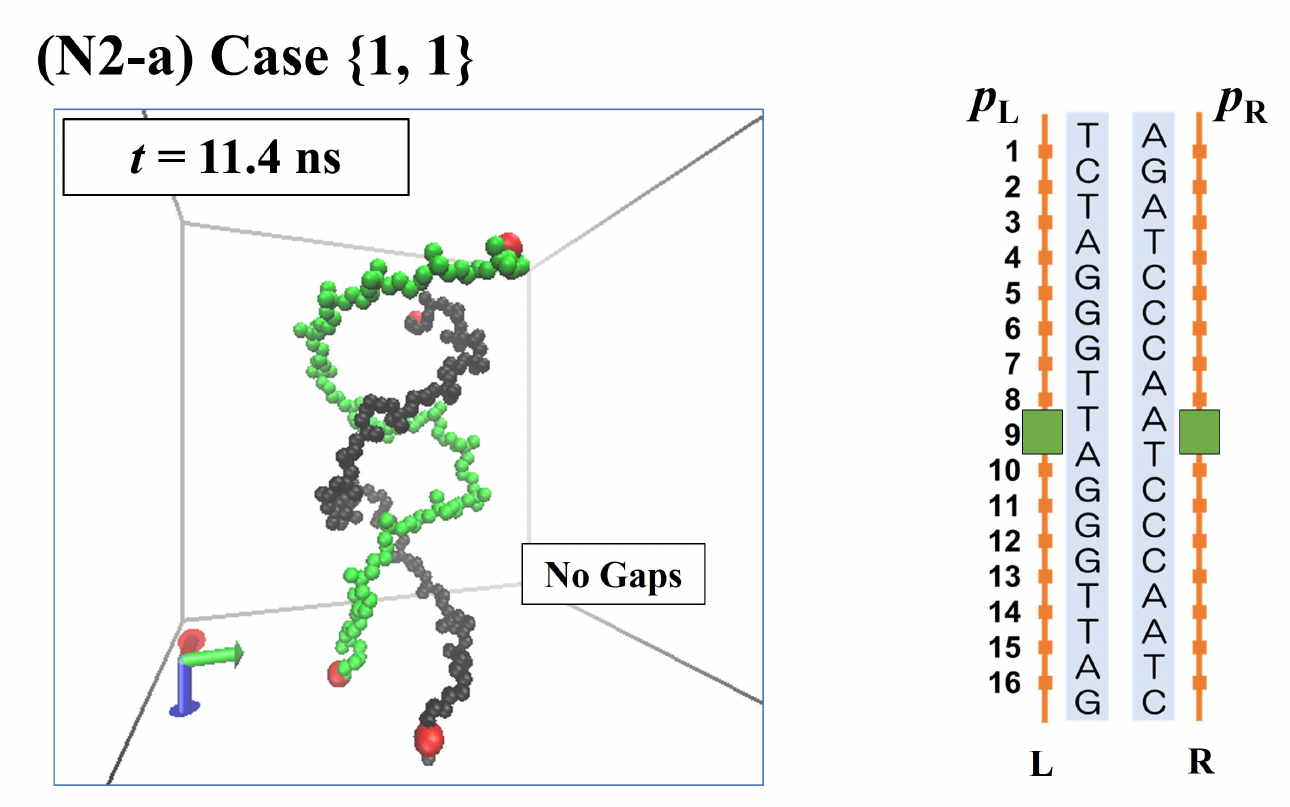} \\
\includegraphics[width=14cm]{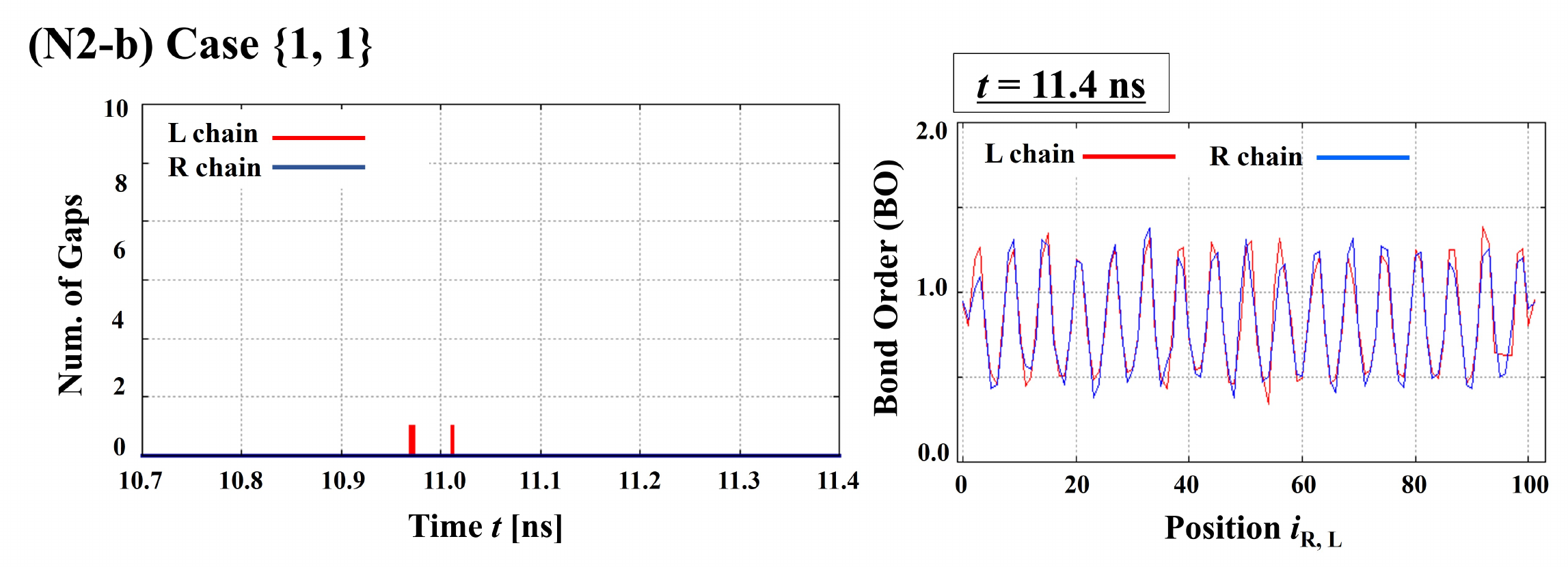}
\caption{
In N2 case $\{1, 1\}, $ there is one scar each on the R and L chains (2 scars in total).
No gaps occur in the R and L chains. 
}
\label{resN2}
\end{figure}

In these cases, \textit{i. e.,} $n_\textrm{L} +  n_\textrm{R}$ = 2 and 8, no gaps appear as shown in Figs. \ref{resN1} and \ref{resN2}. 
In both cases, the structure of the DNA is as stable as the original strands (Fig. \ref{resO}).

\section{Discussions}
DSBs were observed only in calculations with more than 24 scars in the cases of $\left\{16,16\right\}, \left\{16,8\right\}, \left\{8,16\right\}$ and $\left\{12,12\right\}.$ 
Furthermore, SSB occurred for simulations with 16 scars in the cases  $\left\{8,8\right\}, \left\{16,0\right\},$ and $\left\{0,16\right\},$ and no gaps occurred for simulations with less than 16 scars. 
Thus, a significant correlation between the number of scars, and the ease of breaking of strands was observed,

From Table \ref{tab:result} and Figs. \ref{resD1}(D1-b),  \ref{resD2}(D2-b),  \ref{resD3}(D3-b) and \ref{resD4}(D4-b), the number of gaps in DSBs are 8 for D1, 6 for D2, 7 for D3, and 4 for D4, respectively.
Table \ref{tab:result} also shows that SSB, for all three cases (S1, S2, and S3),  occurs at $t \sim 10.9$ ns, which is 0.4 ns earlier than in the original case (O case).
This fact indicates that the installation of scars into the DNA strands increases the fragility of the DNA backbone.
Furthermore, a detailed comparison of the occurrence times from Table \ref{tab:result} reveals that, among the four cases in which DSBs occur, the $\left\{16,16\right\}$ case is the slowest with $t =$ 11.242 ns.
Conversely, the case of $\left\{8,16\right\}$ generates DSBs earliest at 11.041 ns. 
It can, therefore, be concluded that the number of scars alone does not determine the fragility of the strands. 
It is speculated that the occurrence of DSBs in DNA is not solely dependent on the number and location of scars, but also on the solvent effect of ions and water molecules that accumulate around the DNA.

For the cases $\left\{16,0\right\}$ and $\left\{0,16\right\},$ a gap occurs in the scarred strand,  and the other strand does not exhibit any effects resulting from the initial scars.  
In addition, Figs. \ref{resS1}(S1-b) and \ref{resS2}(S2-b) show that the final number of gaps in the $\left\{16,0\right\}$ case is 9 and that in the $\left\{0,16\right\}$ case is 3, indicating a large difference in the number of gaps for the same SSB cases.  
On the other hand, although DSBs occur in the $\left\{16,8\right\}$ and $\left\{8,16\right\}$ cases, the $\left\{16,8\right\}$ case shows a larger number of gaps in the strand with fewer scars. 
In a somewhat unexpected turn of events, in the $\left\{8,16\right\}$ case, the gaps occur at $p_\textrm{L} =$ 15 and 16 on the L strand without any scars.
This indicates that the correlation between the location of the scars and the location of the gaps is not significant.

\begin{figure}[htbp] \centering 
\includegraphics[width=8cm]{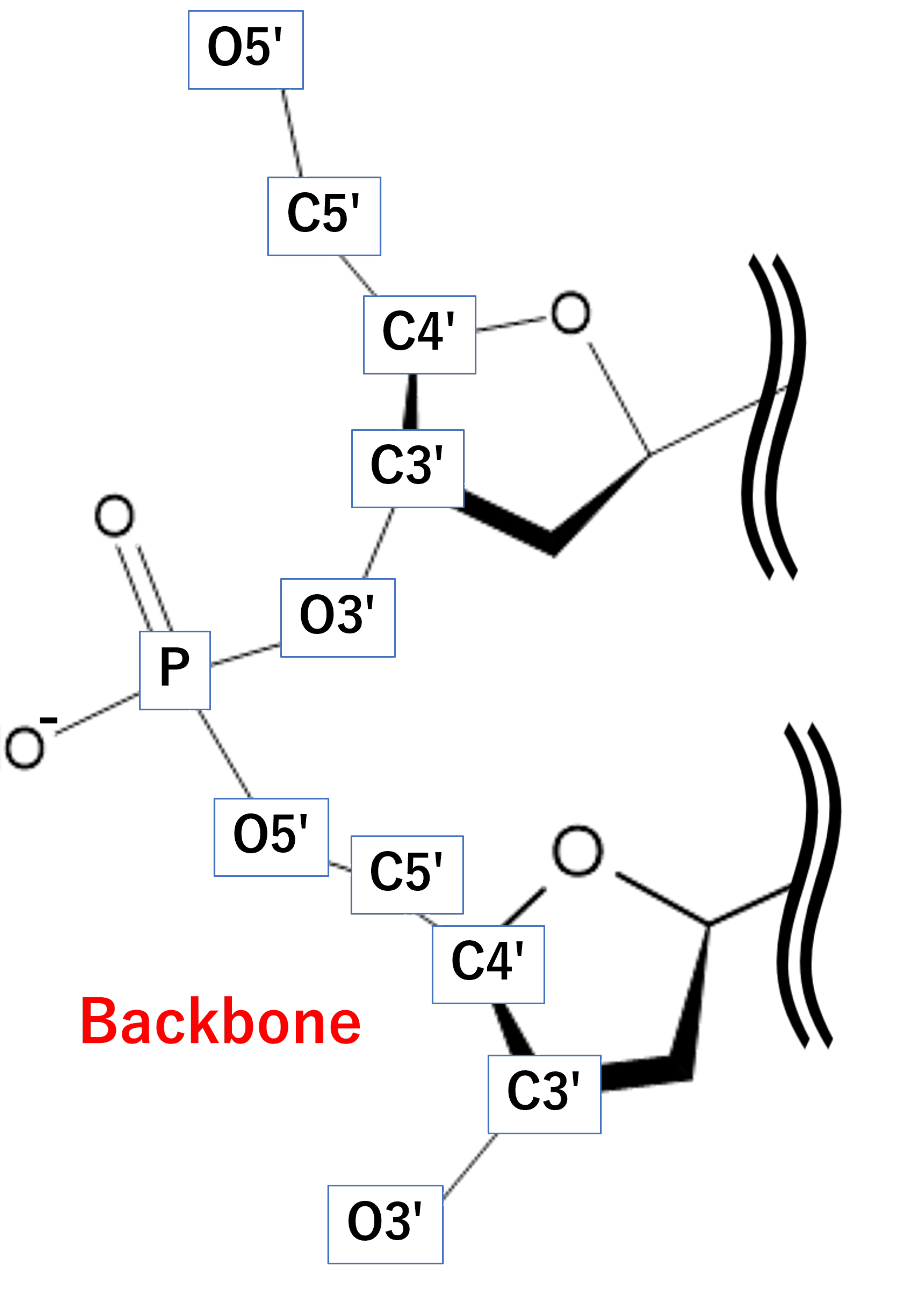}
\caption{The unit structure of telomeric DNA backbones} 
\label{fig.Unit}
\end{figure}

To further investigate the occurrence of gaps in detail, the distribution of gap occurrence at covalent bonds on the backbone of nucleotides (see Fig. \ref{fig.Unit}) is examined.
The locations of the scars that occurred in all 10 simulations in Figs. \ref{fig0060} and \ref{fig0070} are reduced to the `unit' nucleotide.
They are summarised in Table \ref{tab.unit}.
This table shows that the covalent bond between P-O, \textit{i. e.}, the phosphodiester group, is extremely fragile, while the covalent bonds between O-C and C-C around the pentasaccharide residue are resistant to gaps and are durable.
The reason for this can be seen from the fact that the bond order between P-O is only 0.4 to 0.6, even when there is no gap.

\begin{figure}[htbp] \centering 
\includegraphics[width=10cm]{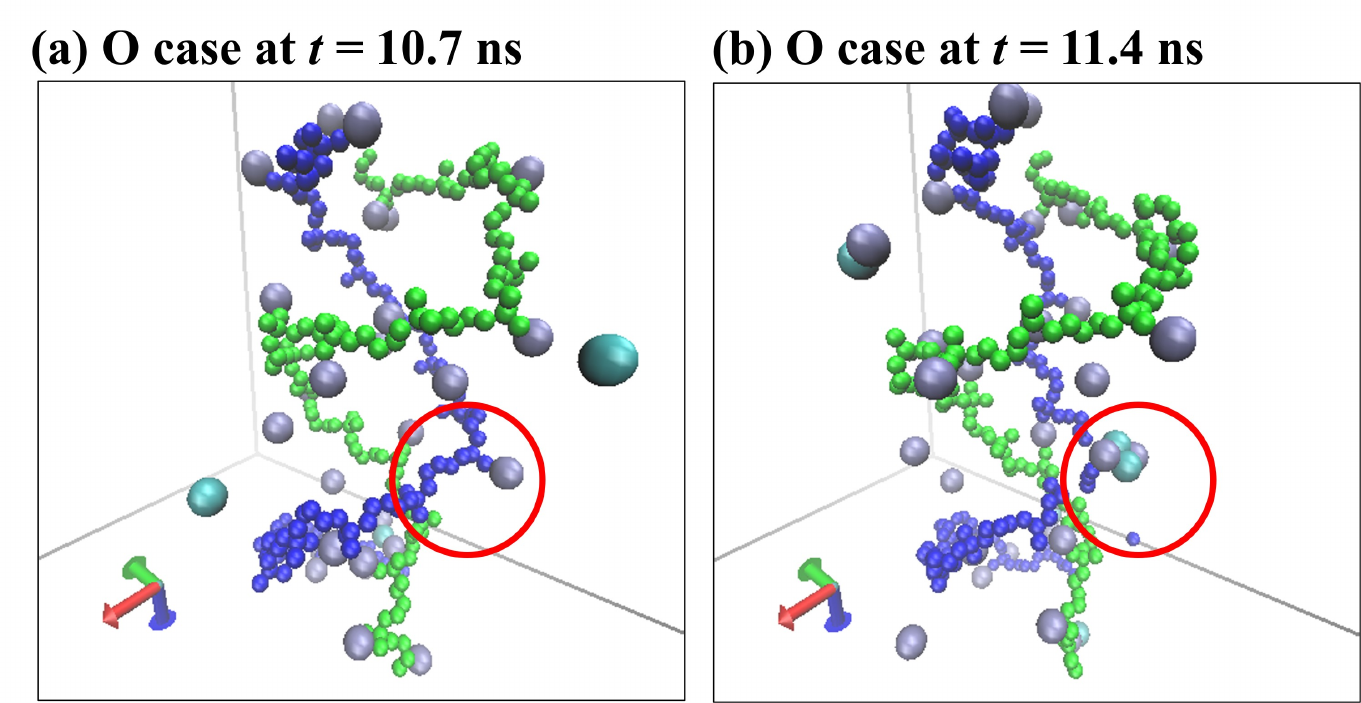}
\caption
{
Sodium (Na$^+$)  and chloride (Cl$^-$) ions aggregate at the gap locations in O case $\{0,0\}.$
A gap occurs at $p_\textrm{L}$ = 7 and 8 circled in red. 
The R or L strand is drawn by green or blue balls, respectively.
Light blue or purple ball means  Na$^+$ or  Cl$^-$, respectively.
The time $t =$ 10.7 ns is in the left-hand figure (a), and 11.12 ns is in the right-hand one (b).
} 
\label{ion}
\end{figure}

Other than the above, a possible cause of the gap in the covalent bond between P-O is the bonding of the phosphate group with ions gathered around it.
Figure \ref{ion} visualizes the ions around the telomeric DNA backbone in the O case $\left\{0,0\right\}.$ 
The left-hand figure shows that Na$^+$ and Cl$^-$  are approaching the phosphate group before the gap occurs at $t=$ 11.12 ns.
Then they interact with P or O in the backbone, weakening the original P-O covalent bond.
Such a phenomenon,  in which multiplet Na$^+$ or Cl$^-$  approach the gap in the backbone, was also observed in other cases.

From the above points, we can analogously conclude that the process that causes the gap is as follows:
\begin{enumerate}
\item{
Excess  Na$^+$ ions are attracted to negatively charged sites (phosphate groups) by hydrogen desorption, or accidental approach by random water movement.
}
\item{
Na$^+$ and phosphate groups make bonds.
}
\item{
The covalent bonds in the backbone are weakened and the phosphate groups are pulled out from the DNA.
}
\end{enumerate}

Thus, it can be said that the more hydrogen desorption and the change of initial charge configuration due to the decomposition effect increase, the more likely the above processes are to occur, and the more the gap increases accordingly. 
This can be inferred from the work of Hishinuma \textit{et al.}\cite{Hishinuma}, where they investigated the process of DNA breaking under irradiation-induced thermal energy using MD simulations.
 According to them, it was reported that when counterions and water molecules are present around the DNA, the phosphate group and Na$^+$ form the intermediate Na$^+$-O-P-O, after which strand breakage is induced at the phosphate group.
This report supports that the processes (1)--(3) described in our study are at work when gaps occur in the DNA.

\begin{table}[htbp] \centering
\caption{Distribution of  covalent bonds on the backbone of the unit nucleotides of telomeric DNA.
In the scar mode, two hydrogens are removed in C5$^\prime$.
}
\label{tab.unit}
\begin{tabular}{||c||c|c||c||}
\Hline
\multicolumn{1}{||c||}{Position} & \multicolumn{3}{c||}{Num. of Gaps} \\ \cline{2-4}
in Backbone &  in L-chain & in R-chain & Total num. \\ \Hline
O3$^\prime$ -- P                     &  5  & 10 &  15  \\ \hline
P -- O5$^\prime$                     & 12 & 5   & 17 \\ \hline
O5$^\prime$ -- C5$^\prime$  &  0  & 3   &  3  \\ \hline
C5$^\prime$ -- C4$^\prime$  &  0  & 3   &  3  \\ \hline
C4$^\prime$ -- C3$^\prime$  &  2  & 0   &  2  \\ \hline
C3$^\prime$ -- O3$^\prime$  &  1  & 0   &  1  \\ \hline
\Hline
\end{tabular}
\end{table}

\section{Conclusion}
We proposed a scar model to reproduce how double-strand breaks occur in the telomeric DNA damaged by the effect of the $\beta$-decay of tritium to helium.
In this scar model, the two hydrogens bonded to the 5$^\prime$ carbon connecting the pentasaccharides and phosphate are removed.
Molecular dynamics simulations using ReaxFF were carried out for 10 cases, with the total number and arrangement of the scars in the strands of the telomeric DNA varying.
The simulation resulted in double-strand breaks being observed for structures with more than 24 scars for the telomeric DNA consisting of 16 base pairs (32 nucleotides). 
In the case of 16 scars, only single-strand breaks were observed.

A more detailed observation of the case of asymmetrical scar distribution on the L and R strands shows that in the case of $\left\{16,0\right\}$ and $\left\{0,16\right\},$ where only one of the strands had scars, single-strand breaks occurred only in the scarred strand and did not grow into double-strand breaks. 
Secondly, in the $\left\{16,8\right\}$ and $\left\{8, 16\right\}$ cases, double-strand breaks occurred and were more pronounced in the strand with fewer scars. 

Based on these results, it can be concluded that the following conditions are necessary for the occurrence of double-strand breaks:
\begin{enumerate}
\item{The ``scars''  must occur on both L and R strands.}
\item{A large number of scars (more than 24) must occur in close proximity to each other.}
\end{enumerate}

\acknowledgment
The computation was performed using Research Center for Computational Science, Okazaki, Japan (Project: 24-IMS-C099) and Plasma Simulator of NIFS. The research was supported by KAKENHI (Nos. 21H04456, 22H05131, 23H04609, 22K18272, 23K03362), by the NINS program of Promoting Research by Networking among Institutions (01422301) by the NIFS Collaborative Research Programs (NIFS22KIIP003, NIFS24KIIT009, NIFS24KIPT013, NIFS22KIGS002, NIFS22KISS021) and by the ExCELLS Special Collaboration Program of Exploratory Research Center on Life and Living Systems(24-S5).

\clearpage

\begin{thebibliography}{99}
\bibitem{02Brenner} D. W. Brenner,O. A. Shenderova, J. A. Harrison, S. J. Stuart, B. Ni and S. B. Sinnott, J. Phys. Condens. Matter \textbf{14}, 783  (2002).
\bibitem{06Ito} A.Ito, H. Nakamura, J. Plasma Phys. \textbf{72}, 805 (2006).
\bibitem{15Ito} A. M. Ito, A. Takayama, Y. Oda, T. Tamura, R. Kobayashi, T. Hattori, S. Ogata, N. Ohno, S. Kajita, M. Yajima, Y. Noiri, Y. Yoshimoto, S. Saito, S. Takamura, T. Murashima, M. Miyamoto and H. Nakamura, J. Nucl. Mat. \textbf{463}, 109 (2015).
\bibitem{11Naka} H. Nakamura, A.M. Ito, S. Saito, A. Takayama, Y. Tmura, N. Ohno, and S. Kajita, Jpn. J. Appl. Phys. \textbf{50},  01AB04 (2011).
\bibitem{23Naka} H. Nakamura, K. Takasan, M. Yajima, S. Saito, J. Adv. Simulat. Sci. Eng. \textbf{10}, 132 (2023).
\bibitem{20Naka} H. Nakamura, H. Miyanishi, T. Yasunaga, S. Fujiwara, T. Mizuguchi, A. Nakata, T. Miyazaki, T. Otsuka, T. Kenmotsu, Y. Hatano and S. Saito, Jpn. J. Appl. Phys. \textbf{59}, SAAE01(2020).
\bibitem{22Hatano} Y. Hatano, H.Nakamura, S. Fujiwara, S. Saito, and T. Kenmotsu, \textit{Damages of DNA in tritiated water}, The Enzymes, Academic Press \textbf{51},  131(2022). 10.1016/bs.enz.2022.08.009. 
\bibitem{19Fujiwara}S. Fujiwara, H. Nakamura, H. Li, H. Miyanishi, T. Mizuguchi, T. Yasunaga, T. Otsuka, Y. Hatano and S. Saito, J. Adv. Simulat. Sci. Eng. \textbf{6}, 94 (2019). 
\bibitem{19Li} H. Li, S. Fujiwara, H. Nakamura, T. Mizuguchi, T. Yasunaga, A. Nakata, T. Miyazaki, T. Otsuka, T. Kenmotsu, Y. Hatano, S.Saito, Plasma Fus. Res. \textbf{14}, 3401106 (2019). 
\bibitem{Allen}   M. P. Allen and D. J. Tildesley, \textit{Computer Simulation of Liquids} (Oxford University Press, New York, 1991).
\bibitem{Frenkel}   D. Frenkel and B. Smit, \textit{Understanding Molecular Simulations: From Algorithms to Applications} (Academic, San Diego, 2002).
\bibitem{17Tanabe} T. Tanabe (ed.), \textit{Tritium: Fuel of Fusion Reactors} (Springer Japan, Japan, 2017).
\bibitem{16Tritium} United Nations Scientific Committee on the Effects of Atomic Radiation, \textit{Sources, Effects and Risks of Ionizing Radiation Annex C Biological Effects of Selected Internal Emitters-Tritium} (United Nations Publication 2016).
\bibitem{09Mull} L. Mullenders, M. Atkinson, H. Paretzke, L. Sabatier, S. Bouffler, Nat. Rev. Cancer \textbf{9}, 596 (2009).
\bibitem{18Hatano}Y. Hatano, Y. Konaka, H. Shimoyachi, T. Kenmotsu, Y. Oya, H. Nakamura, Fusion Eng. Des., \textbf{146}, 1200 (2018). 
\bibitem{CHARMM36DNA} K. Hart, N. Foloppe, C. M. Baker, E. J. Denning, L. Nilsson, A. D. MacKerell Jr., J. Chem. Theory Comput., \textbf{8}, 348 (2012).
\bibitem{99Greider} C. W. Greider, Cell \textbf{97}, 419 (1999).
\bibitem{3SJM} S.K. Nair, S.K. Sliverman, J.H. Chen, and Y. Xiao, Crystal structure analysis of TRF2-Dbd-DNA complex. 2012, \texttt{http://dx.doi.org/10.2210/pdb3SJM/pdb}.
\bibitem{CHARMMpot} B. R. Brooks, C. L. Brooks III, A. D. Mackerell Jr, L. Nilsson, R. J. Petrella, B. Roux, Y. Won, G. Archontis, C. Bartles, S. Boresch, A. Caflish, L. Caves, Q. Cui, A. R. Dinner, M. Feig, S. Fischer, J. Gao, M. Hodoscek, W. Im, K. Kuczera, T. Lazaridis, J. Ma, V. Ovchinnikov, E. Paci, R. W. Pastor, C. B. Post, J. Z. Pu, M. Schaefer, B. Tidor, R. M. Venable, H. L. Woodcock, X. Wu, W. Yang, D. M. York, M. Karplus, J. Comp. Chem. \textbf{30}, 1545 (2009).
\bibitem{LAMMPS} A. P. Thompson, H. M. Aktulga, R. Berger, D. S. Bolintineanu, W. M. Brown, P. S. Crozier, P. J. in 't Veld, A. Kohlmeyer, S. G. Moore, T. D. Nguyen, R. Shan, M. J. Stevens, J. Tranchida, C. Trott, S. J. Plimpton, Comp. Phys. Comm. \textbf{271}, 10817 (2022). 
\bibitem{LAMMPSURL} https://www.lammps.org/index.html.
\bibitem{Langevin} T. Schneider and E. Stoll, Phy. Rev. B. \textbf{17}, 1302 (1978).
\bibitem{v2} S. Monti, A. Corozzi, P. Fristrup, K. L. Joshi, Y. K. Shin, P. Oelschlaeger, A. C. T. van Duin and V. Barone,, Phys. Chem. Chem. Phys. \textbf{51}, 15062 (2013).
\bibitem{ReaxFF} H. M. Aktulga, J. C. Fogarty, S. A. Pandit, and A. Y. Grama, Parallel Computing, \textbf{38}, 245 (2012).
\bibitem{Reax18} A. Vashisth, C.Ashraf, W. Zhang, C.E. Bakis, and A.C. T. van Duin, J. Phys. Chem. A, \textbf{122}, 6633 (2018).
\bibitem{Tsuchida} Y. Tsuchida, S, Saito, H. Nakamura, Y. Yonetani, S. Fujiwara, Nihon Simulation Gakkai Ronbunshi, \textbf{13}, 32 (2021) in Japanese.
\bibitem{Gaussian} Gaussian 09, Revision D.01, M. J. Frisch, G. W. Trucks, H. B. Schlegel, G. E. Scuseria, M. A. Robb, J. R. Cheeseman, G. Scalmani, V. Barone, G. A. Petersson, H. Nakatsuji, X. Li, M. Caricato, A. Marenich, J. Bloino, B. G. Janesko, R. Gomperts, B. Mennucci, H. P. Hratchian, J. V. Ortiz, A. F. Izmaylov, J. L. Sonnenberg, D. Williams-Young, F. Ding, F. Lipparini, F. Egidi, J. Goings, B. Peng, A. Petrone, T. Henderson, D. Ranasinghe, V. G. Zakrzewski, J. Gao, N. Rega, G. Zheng, W. Liang, M. Hada, M. Ehara, K. Toyota, R. Fukuda, J. Hasegawa, M. Ishida, T. Nakajima, Y. Honda, O. Kitao, H. Nakai, T. Vreven, K. Throssell, J. A. Montgomery, Jr., J. E. Peralta, F. Ogliaro, M. Bearpark, J. J. Heyd, E. Brothers, K. N. Kudin, V. N. Staroverov, T. Keith, R. Kobayashi, J. Normand, K. Raghavachari, A. Rendell, J. C. Burant, S. S. Iyengar, J. Tomasi, M. Cossi, J. M. Millam, M. Klene, C. Adamo, R. Cammi, J. W. Ochterski, R. L. Martin, K. Morokuma, O. Farkas, J. B. Foresman, and D. J. Fox, Gaussian, Inc., Wallingford CT, 2016.
\bibitem{B3} A. D. Becke, J. Chem. Phys., \textbf{98}, 5648(1993).
\bibitem{LYP} C. Lee, W. Yang, R. G. Parr, Phys. Rev. B \textbf{37}, 785(1988).
\bibitem{ccpvdz} T. H. Dunning, Jr., J. Chem. Phys. \textbf{90}, 1007(1989).
\bibitem{bond} T.P. Senftle, S.Hong, M. M. Islam, S. B. Kylasa, Y. Zheng, Y. K. Shin, C. Junkermeier, R. Engel-Herbert, M. J. Janik, H. M. Aktulga, T. Verstraelen, A.Grama and A.C.T.  van Duin, npj, Comput. Mater. \textbf{2}, 15011(2016).
\bibitem{Hishinuma} N. Hishinuma, K. Oikawa, T. Okada, M. Kanno, H. Yamazaki, W. C. Chung, A. Saito, H. Kohno, SENAC, \textbf{50}, 3 (2017) in Japanese.
\end{thebibliography}
\end{document}